%% file: paper.tex
\documentclass[aps, pre, a4paper, floatfix,  twocolumn, longbibliography, nofootinbib]{revtex4-1}

\usepackage[utf8]{inputenc}
\usepackage{epsfig}
\usepackage[T1]{fontenc}
\usepackage[english]{babel}
\usepackage[table]{xcolor}
\usepackage{t1enc}
\usepackage{graphicx}
\usepackage{amssymb}
\usepackage{amsmath}
\usepackage{relsize}
\usepackage[percent]{overpic}
\usepackage{bm}
\usepackage{hyperref}

\usepackage{siunitx}
\usepackage{accents}
\usepackage{mathtools}

\usepackage{grffile}

\usepackage[normalem]{ulem}

\usepackage{pgf}
\usepackage{tikz}

\newcommand{\avg}[1]{{\left<#1\right>}}

\newcommand{\dd}{\mathrm{d}}

\usepackage{mathtools}
\def\multiset#1#2{\ensuremath{\left(\kern-.3em\left(\genfrac{}{}{0pt}{}{#1}{#2}\right)\kern-.3em\right)}}

\usepackage{amsmath}

\newcommand{\A}{\bm{A}}
\newcommand{\G}{\bm{G}}
\newcommand{\g}{\bm{g}}
\newcommand{\bb}{\bm{b}}

\newcommand{\E}{\mathcal{E}}
\newcommand{\M}{\mathcal{M}}
\newcommand{\X}{\mathcal{X}}
\newcommand{\T}{\mathcal{T}}

\newcommand{\etal}{\textit{et al}}

\newcolumntype{P}[1]{>{\centering\arraybackslash}p{#1}}

\usepackage{verbatim}
\usepackage{overpic}
\usepackage{booktabs}
\usepackage{placeins}

\usepackage{siunitx}
\usepackage{multirow}

\begin{document}

\title{Disentangling homophily, community structure and triadic closure in networks}

\author{Tiago P. Peixoto}
\email{peixotot@ceu.edu}
\affiliation{Department of Network and Data Science, Central European University, 1100 Vienna, Austria}
\affiliation{Department of Mathematical Sciences, University of Bath, Claverton Down, Bath BA2
  7AY, United Kingdom}

\begin{abstract}
  Network homophily, the tendency of similar nodes to be connected, and
  transitivity, the tendency of two nodes being connected if they share
  a common neighbor, are conflated properties in network analysis, since
  one mechanism can drive the other. Here we present a generative model
  and corresponding inference procedure that are capable of
  distinguishing between both mechanisms. Our approach is based on a
  variation of the stochastic block model (SBM) with the addition of
  triadic closure edges, and its inference can identify the most
  plausible mechanism responsible for the existence of every edge in the
  network, in addition to the underlying community structure itself. We
  show how the method can evade the detection of spurious communities
  caused solely by the formation of triangles in the network, and how it
  can improve the performance of edge prediction when compared to the
  pure version of the SBM without triadic closure.
\end{abstract}

\maketitle
%\tableofcontents

\section{Introduction}

One of the most typical properties of social networks is the presence of
\emph{homophily}~\cite{mcpherson_homophily_1987,shrum_friendship_1988,moody_race_2001,mcpherson_birds_2001},
i.e. the increased tendency of an edge to exist between two nodes if
they share the same underlying characteristic, such as race, gender,
class and a variety of other social parameters. More broadly, when the
underlying similarity parameter is not specified \emph{a priori}, the
same homophily pattern is known as \emph{community
structure}~\cite{fortunato_community_2010}.  Another pervasive pattern
encountered in the same kinds of network is
\emph{transitivity}~\cite{rapoport_spread_1953,holland_transitivity_1971,holland_local_1975},
i.e. the increased probability of observing an edge between two nodes if
they have a neighbor in common. Although these patterns are indicative
of two distinct mechanisms of network formation, namely choice or
constraint homophily~\cite{kossinets_origins_2009} and triadic
closure~\cite{granovetter_strength_1973}, respectively, they are
difficult to distinguish in non-longitudinal data. This is because both
processes can result in the same kinds of observation: 1. the preferred
connection between nodes of the same kind can induce the presence of
triangles involving similar nodes, and 2. the tendency of triangles to
be formed can induce the formation of groups of nodes with a higher
density of connections between them, when compared to the rest of the
network~\cite{foster_clustering_2011,foster_communities_2010}. This
conflation means we cannot reliably interpret the underlying mechanisms
of network formation merely from the abundance of triangles or observed
community structure in network data.

In this work we present a solution to this problem, consisting in a
principled method to disentangle homophily and community structure from
triadic closure in network data, conditioned on mild modeling
assumptions. This is achieved by formulating a generative model that
includes community structure in a first instance, and an iterated
process of triadic closure in a second. Based on this model, we develop
a nonparametric Bayesian inference algorithm that is capable of
identifying which edges are more likely to be due to community structure
or triadic closure, in addition to the underlying community structure
itself. What our approach demonstrates is that, while at first it seems
that triadic closure and homophily generate similar patterns in network
structure, the different mechanisms also leave behind particular traces
in the network structure that can be used to disambiguate between the
two.

Several authors have demonstrated that triadic closure can induce
community structure and homophily in networks. Foster
\etal~\cite{foster_clustering_2011,foster_communities_2010} have shown
that maximum entropy network ensembles conditioned on prescribed
abundances of triangles tend to possess high modularity. A more recent
analysis of this kind of ensemble by López
\etal~\cite{lopez_transitions_2020} showed that it is marked by a
spontaneous size-dependent formation of ``triangle clusters.'' Bianconi
\etal~\cite{bianconi_triadic_2014} have investigated a network growth
model, where nodes are progressively added to the network, and connected
in such a way as to increase the amount of triangles, and shown that it
is capable of producing networks with emergent community structure. The
effect of triangle formation on apparent community structure has been
further studied by Wharrie \etal~\cite{wharrie_micro-_2019}, who showed
that those patterns can even mislead methods specifically designed to
avoid the detection of spurious communities in random networks. More
recently, Asikainen \etal~\cite{asikainen_cumulative_2020} have shown
that iterated triadic closure can exacerbate homophily present in the
original network, via a simple macroscopic model.

The approach presented in this work differs from the aforementioned ones
primarily in that it runs in the reverse direction: instead of only
defining a conceptual network model that demonstrates the interlink
between triadic closure and homophily given prescribed parameters, the
proposed method operates on empirical network data, and reconstructs the
underlying generative process, decomposing it into distinct community
structure and triadic closure components. As we show, this
reconstruction yields a detailed interpretation of the underlying
mechanisms of network formation, allowing us to identify macro-scale
structures that emerge spontaneously from micro-scale higher-order
interactions~\cite{battiston_networks_2020,benson_simplicial_2018}, and
in this way we can separate them from inherently macro-scale structures.

It is also worth mentioning some recent methods that have been proposed
that use triangles as a means of finding communities in
networks~\cite{palla_uncovering_2005,benson_higher-order_2016,yin_local_2017}. Although
these methods can be informative of the interplay between triangles and
large-scale structure, they cannot explain the formation of the
triangles themselves, or identify the contribution of pairwise
homophily, as we do here. Likewise, there are also methods that
reconstruct networks via compositions of higher-order building
blocks~\cite{wegner_atomic_2021,young_hypergraph_2021}, but which can
make no statement about any existing large-scale homophily. Finally, a
commonly used approach in the social sciences literature is to model the
occurrence of triangles and homophily using exponential random graph
models (ERGMs)~\cite{robins_introduction_2007}. Generally, these models
do not possess likelihoods that can be expressed in closed form, making
their inference quite difficult without relying on
approximations. Furthermore, when they are used to model the presence of
triangles or other small subgraphs, they tend to possess extreme
degeneracies~\cite{strauss_model_1975, park_statistical_2004,
park_solution_2004, foster_clustering_2011, fischer_sampling_2015},
rendering them rather implausible models for clustered
networks. Additionally, when they are combined with homophily, this only
usually done with observed homophilic traits, not latent ones as we
consider here.

Our method is based on the nonparametric Bayesian inference of a
modified version of the stochastic block model
(SBM)~\cite{holland_stochastic_1983,peixoto_bayesian_2019} with the
addition of triadic closure edges, and therefore leverages the
statistical evidence available in the data, without
overfitting. Importantly, our method is capable of determining when the
observed structure can be attributed to an actual preference of
connection between nodes, as described by the SBM, rather than an
iterated triadic closure process occurring on top of a substrate
network. As a result, we can distinguish between ``true'' and
\emph{apparent} community structure caused by increased transitivity.  A
key concept in the method that allows this distinction to be made is the
principle of maximum parsimony: in situations where both transitivity
and homophily serve as competing hypotheses, their relative plausibility
is evaluated based not only on how well they can explain the data, but
also on the amount of information needed to specify the particular model
in the first place. As we also demonstrate, this decomposition yields an
edge prediction method that tends to perform better in many instances
than the SBM used in isolation.

We emphasize that our approach is capable of performing the
decomposition between homophily and triadic closure from a single
network observation without annotations. At first, this might seem at
odds with formal results relating to similar, but distinct decomposition
problems, that state that this kind of disentanglement is not possible
from a single network observation. In particular, Chang et
al~\cite{chang_estimation_2020} considered a scenario of uncertain
network measurement, and proved that, absent of any modeling assumption
on how the edges of the network are initially placed, it is not possible
to estimate the network structure from a single network
observation. Similarly, Shalizi and Thomas~\cite{shalizi_homophily_2011}
famously proved that contagion (causal inheritance of traits due to peer
influence) cannot be distinguished from homophily given a single network
observation. Both of these statements rely on a lack of stipulation on
how the networks are generated (which formally cannot be distinguished
from making an explicit assumption that all networks are equally likely
\emph{a priori}). However, whenever such stipulations are made, the
situation changes. In particular, McFowland III and
Shalizi~\cite{mcfowland_iii_estimating_2021} have shown that as soon as
the homophilic traits are latent (instead of being observed directly as
considered in Ref.~\cite{shalizi_homophily_2011}), and can be modeled as
a SBM, the disentanglement becomes possible, even for a single
network. Likewise, if we use the SBM as a structured prior
distribution~\cite{peixoto_reconstructing_2018}, it becomes possible to
estimate the magnitude of the measurement error as well as to
reconstruct noisy networks, even for a single network observation and
when the error magnitude is unknown \emph{a priori}. Although the
disentanglement problem that we consider here is different from the
aforementioned ones, and the impossibility results do not carry over, we
nevertheless make use of the same kinds of modeling assumptions that
make the other problems feasible.

Our manuscript is organized as follows. In Sec.~\ref{sec:model} we
describe our model, and its inference procedure. In
Sec.~\ref{sec:artificial} we demonstrate how it can be used disambiguate
triadic closure from community structure in artificially generated
networks. In Sec.~\ref{sec:empirical} we perform an analysis of
empirical networks, in view of our method. In Sec.~\ref{sec:prediction}
we show how our model can improve edge prediction. We end in
Sec.~\ref{sec:conclusion} with a conclusion.

\section{Stochastic block model with triadic closure (SBM/TC)}\label{sec:model}

\begin{figure}
  \resizebox{\columnwidth}{!}{
  \begin{tabular}{cccc}
    \multicolumn{4}{c}{\larger Generative process}\\\hline\\[-.7em]
    Seminal edges & \multicolumn{2}{c}{Triadic closure} & Observed network\\
    \includegraphics[width=.33\columnwidth]{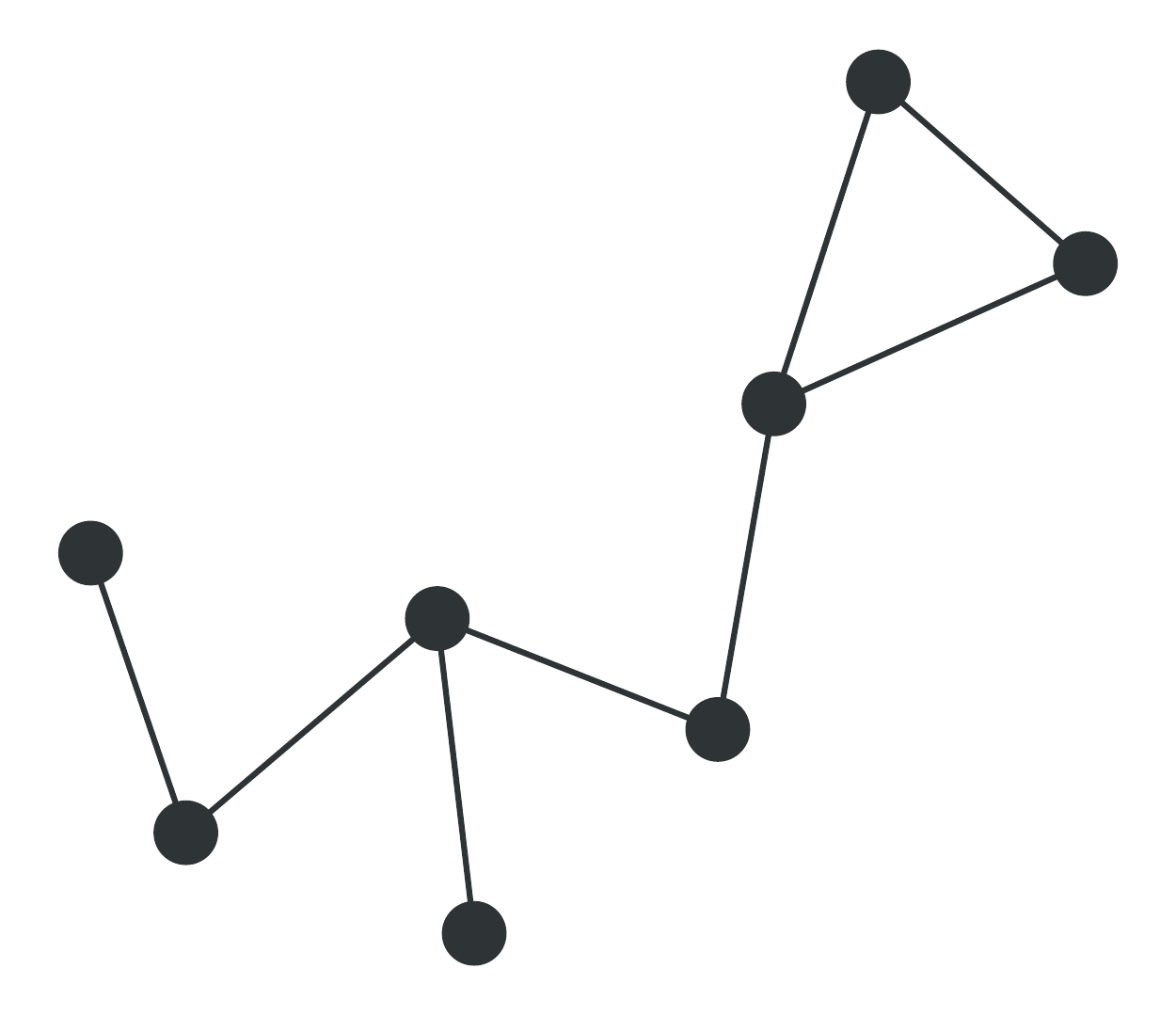}&
    \multicolumn{2}{c}{\includegraphics[width=.33\columnwidth]{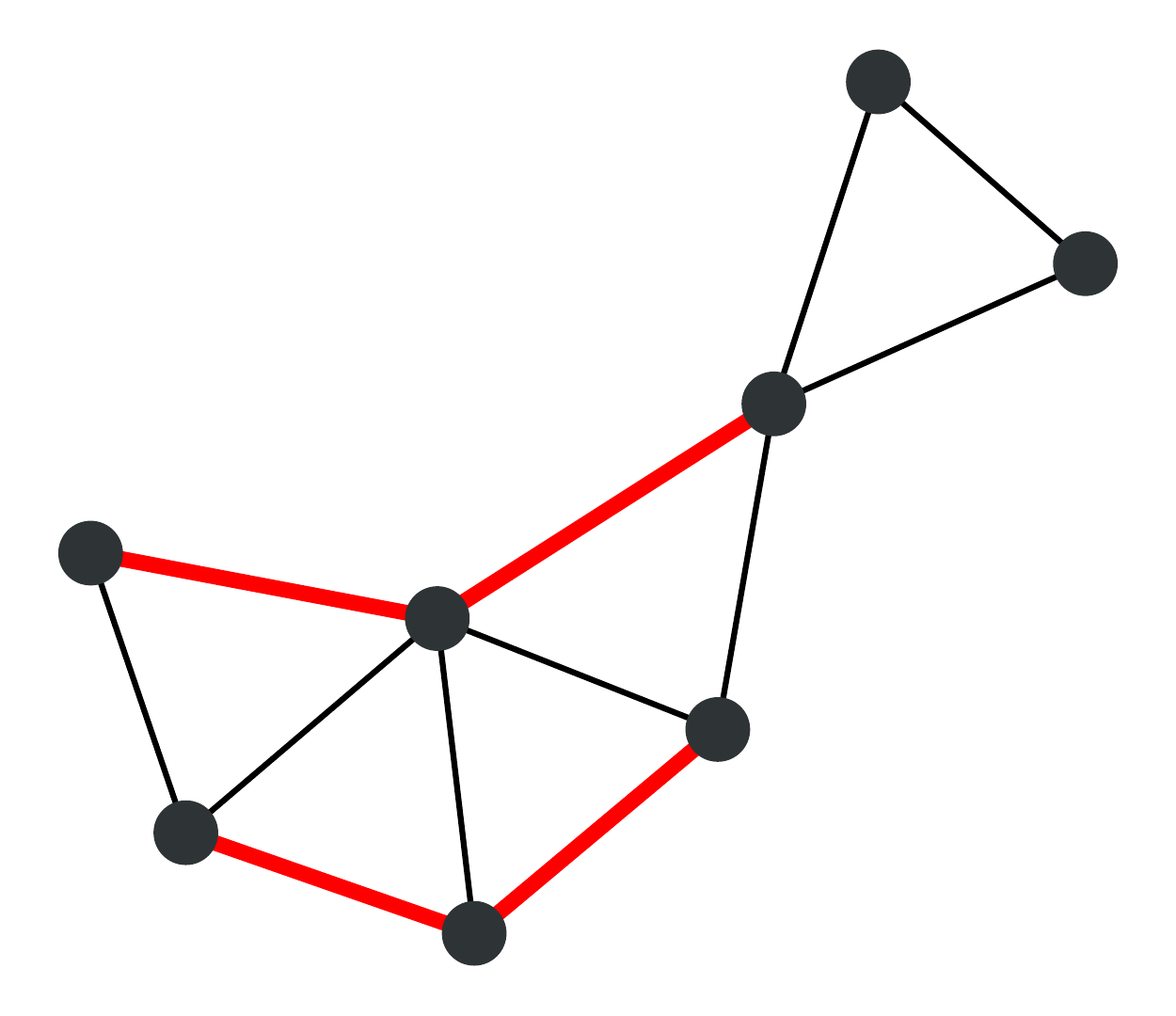}}&
    \includegraphics[width=.33\columnwidth]{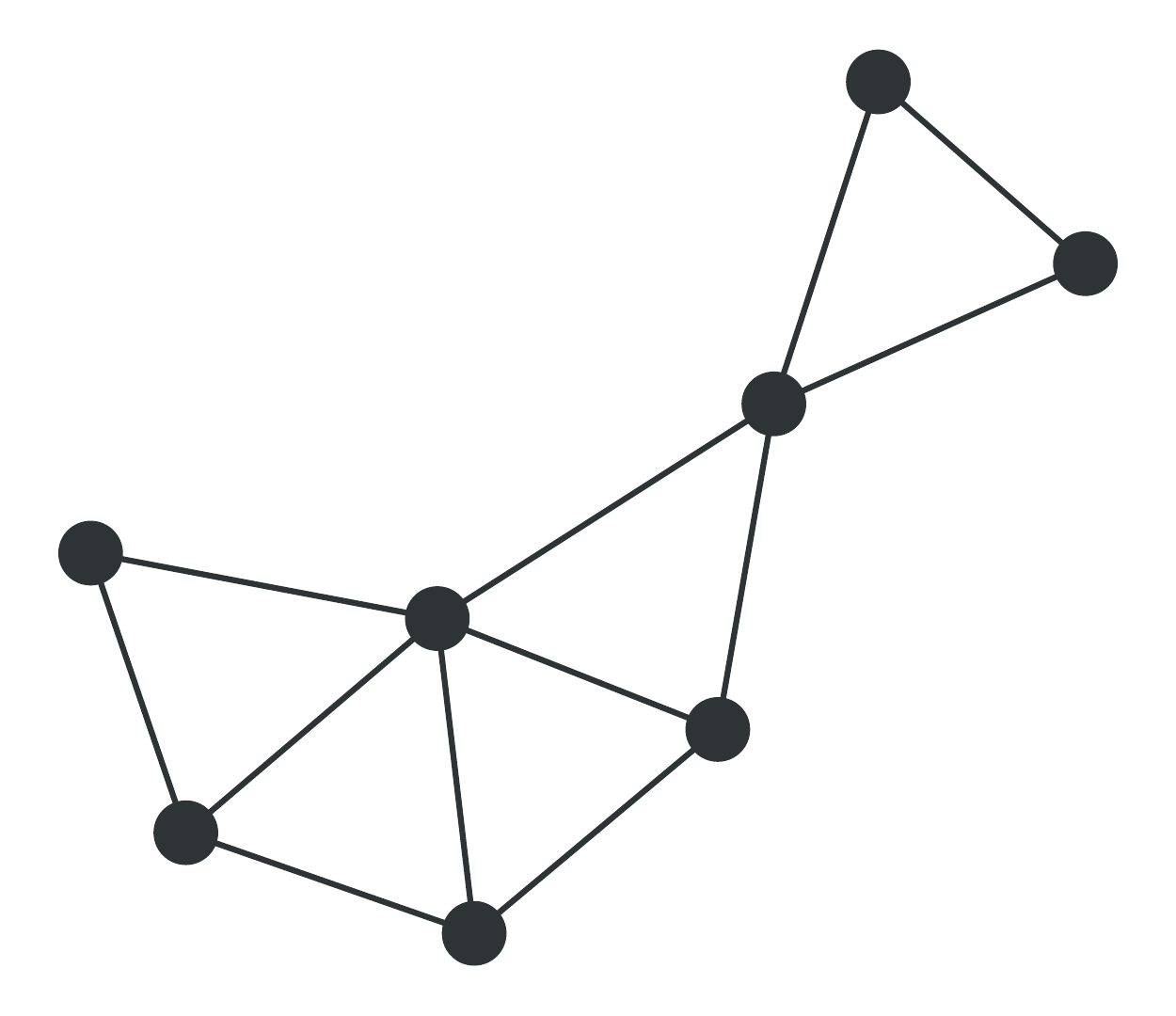}\\
    \multicolumn{4}{c}{\larger Statistical inference}\\\hline\\[-.7em]
    Observed network & \multicolumn{2}{c}{Posterior distribution} & Marginal probabilities\\[.5em]
    \multirow{2}{*}[1cm]{\includegraphics[width=.33\columnwidth]{diagram.pdf}}&
    \includegraphics[width=.15\columnwidth]{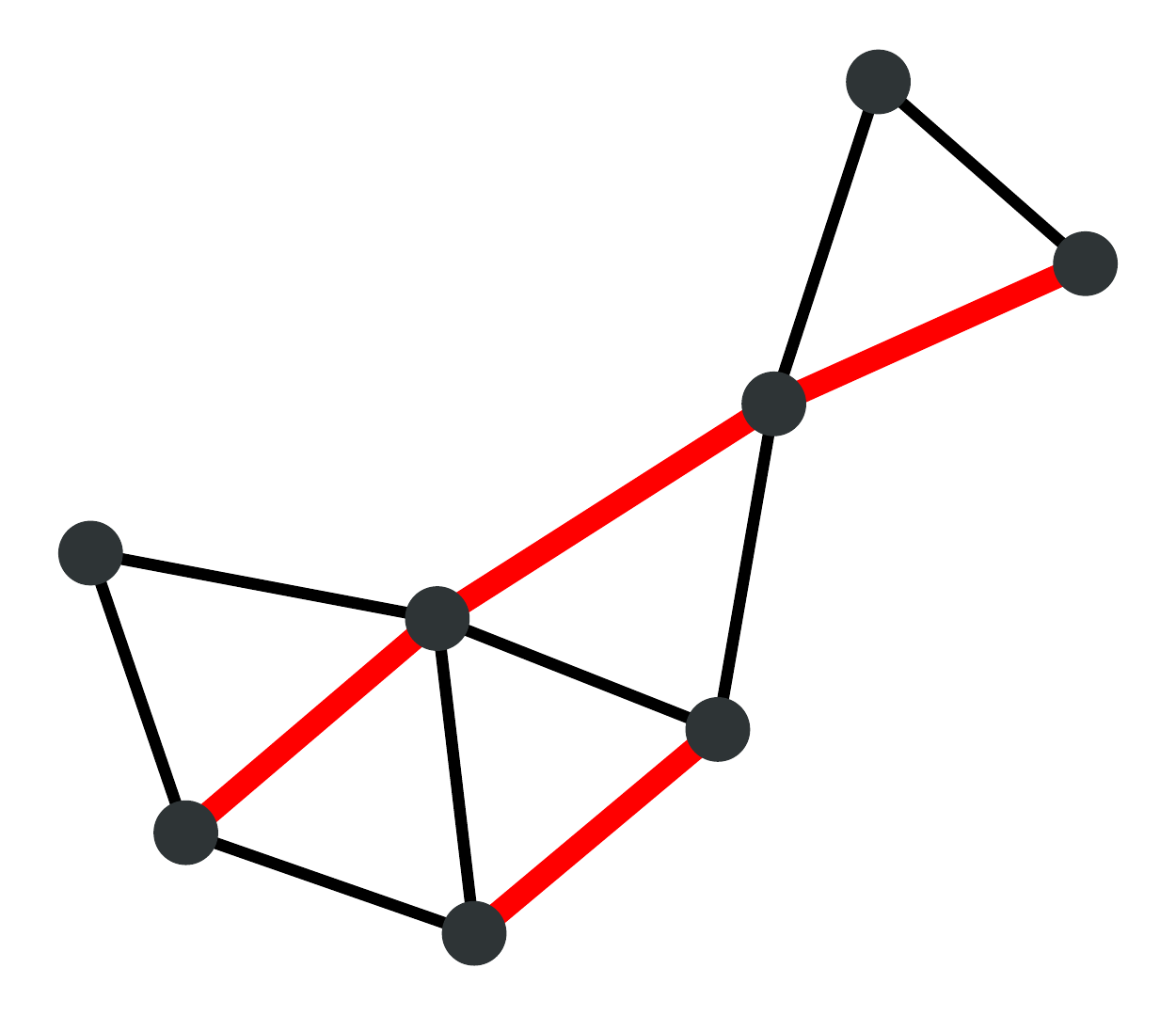}&
    \includegraphics[width=.15\columnwidth]{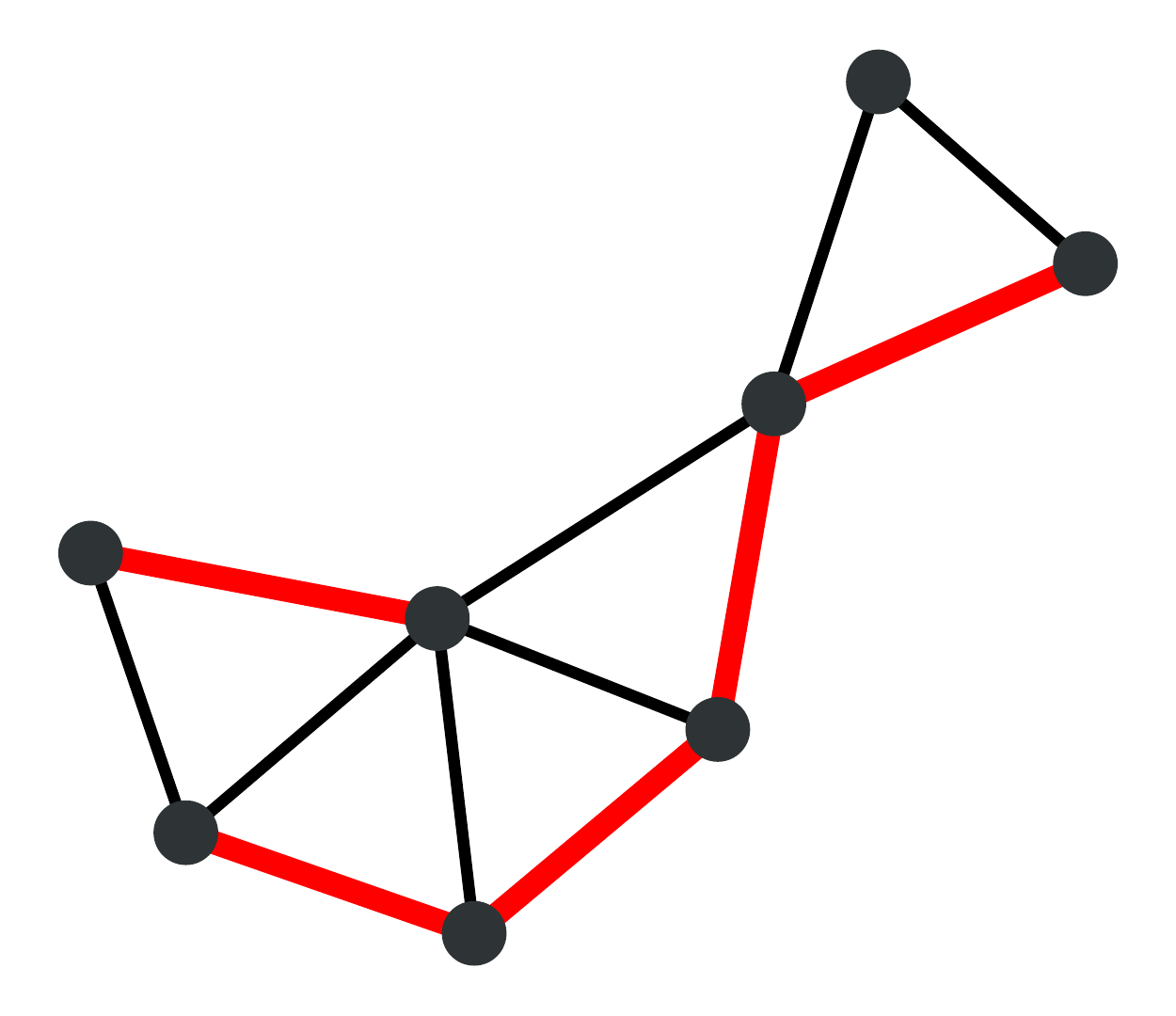}&
    \multirow{2}{*}[1cm]{\includegraphics[width=.33\columnwidth]{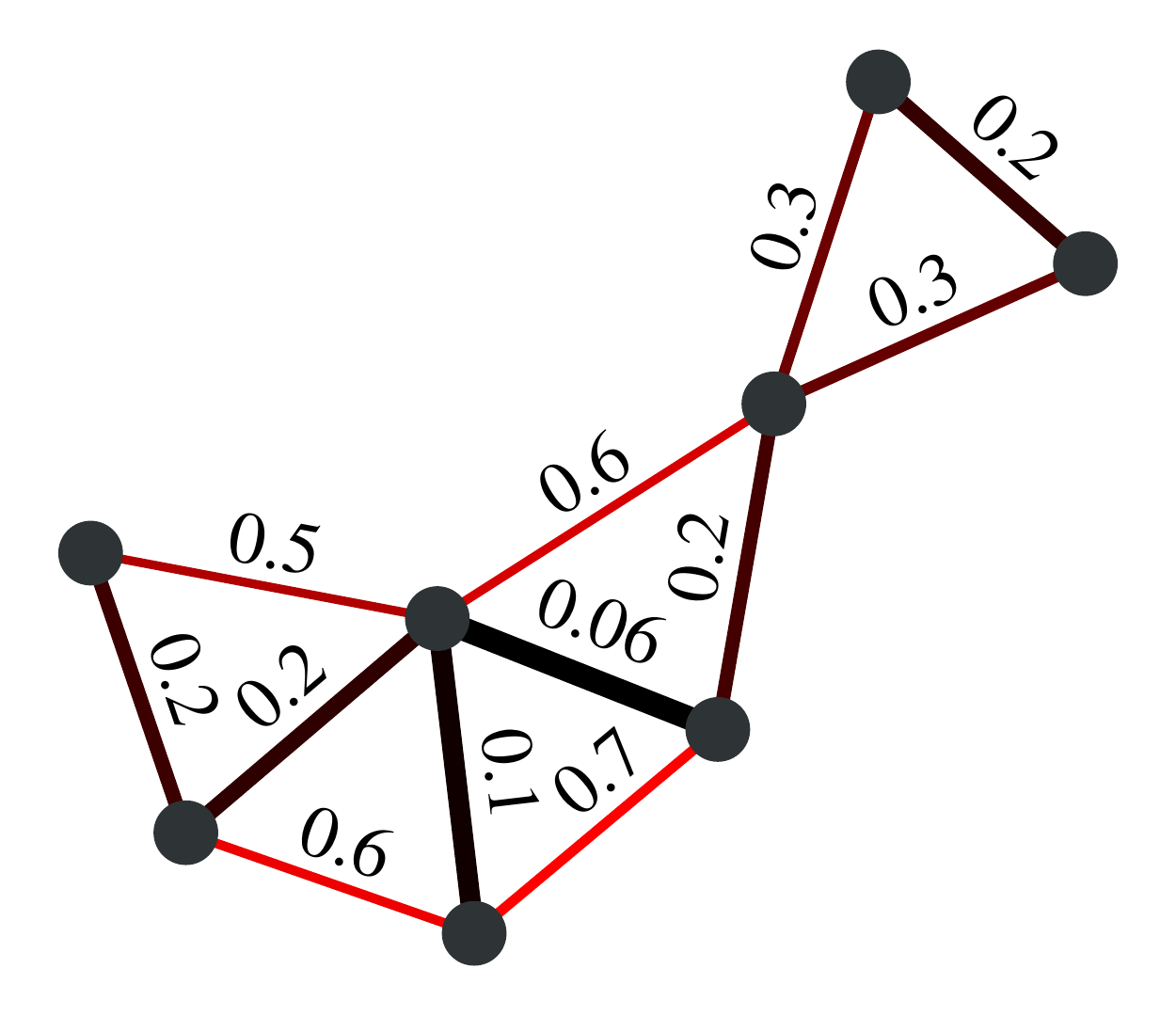}}\\
    &\includegraphics[width=.15\columnwidth]{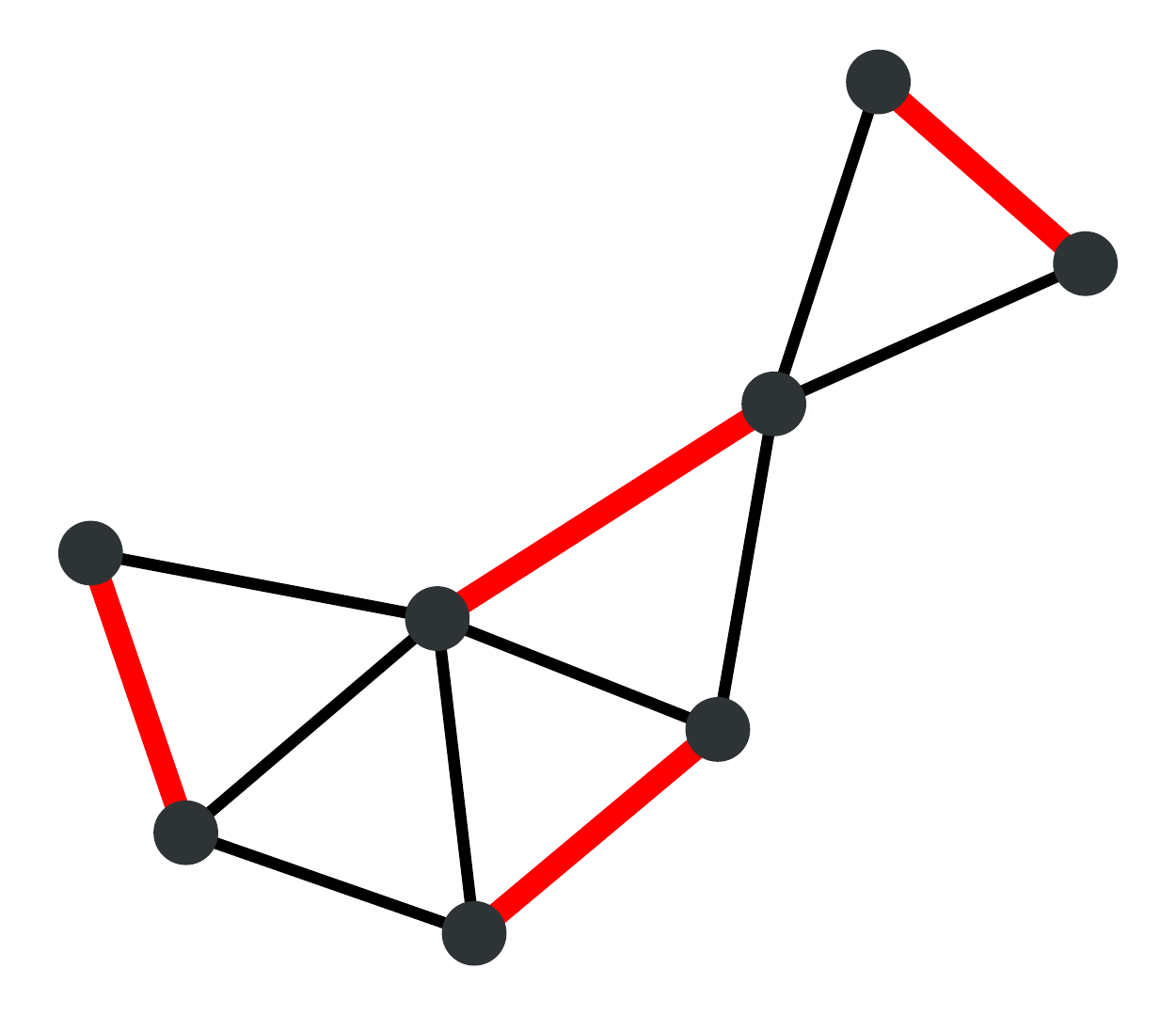} &
    \includegraphics[width=.15\columnwidth]{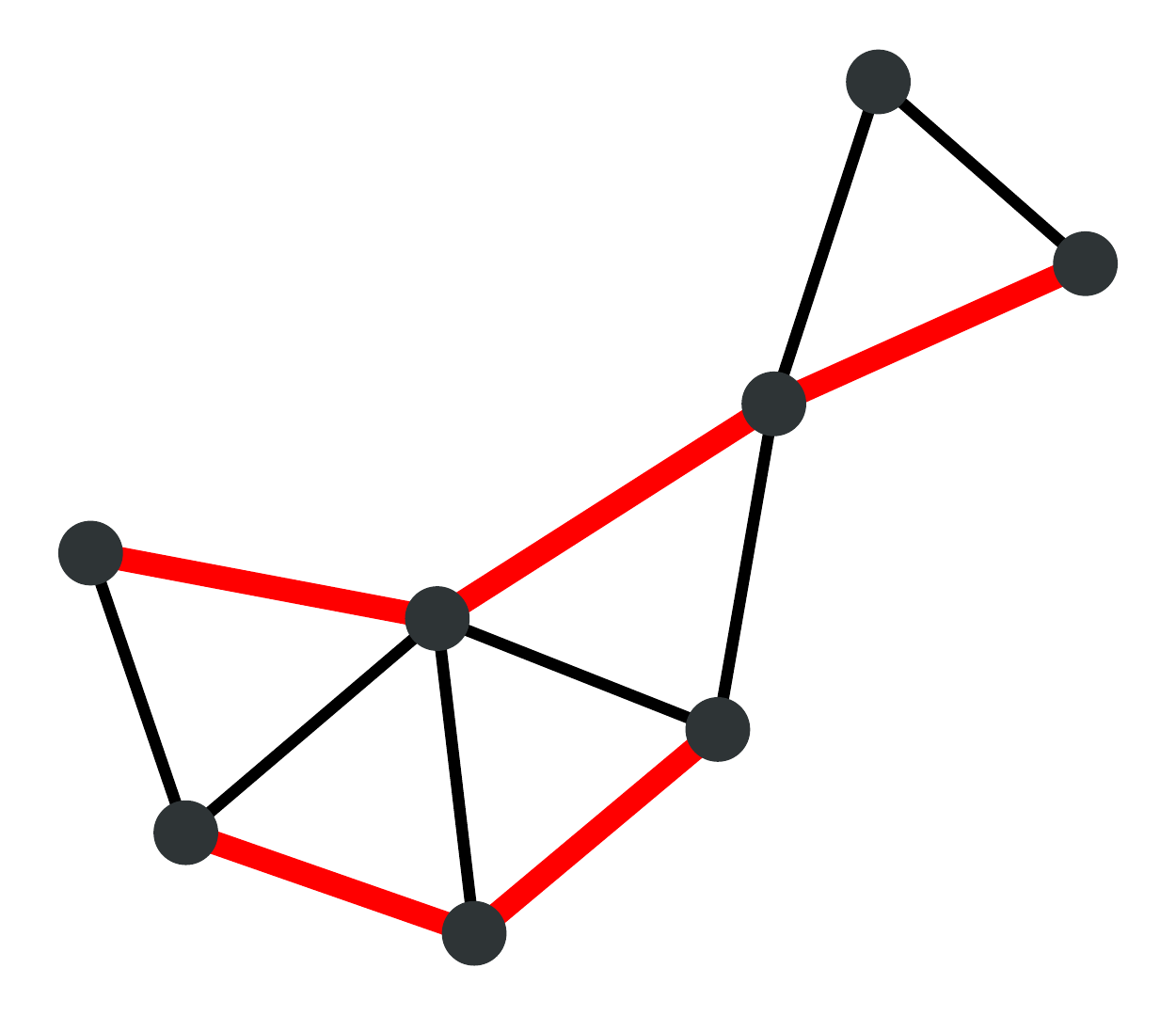} &
  \end{tabular}} \caption{Schematic representation of the generative
  process considered (top) and the associated inference procedure
  (bottom). The generative process consists in the placement of seminal
  edges according to a SBM, and the addition of triadic closure edges
  conditioned on the seminal edges (shown in red). The inference
  procedure runs in the reverse direction, and given an observed graph,
  it produces a posterior distribution of possible divisions of seminal
  and triadic closure edges, with which edge marginal probabilities on
  the edge identities can be obtained.\label{fig:diagram}}
\end{figure}

Community structure and triadic closure are generally interpreted as
different processes of network formation. With the objective of allowing
their identification \emph{a posteriori} from network data, our approach
consists in defining a generative network model that encodes both
processes explicitly. More specifically, our generative model consists
of two steps, with the first one being the generation of a substrate
network containing ``seminal'' edges, placed according to an arbitrary
mixing pattern between nodes, and an additional layer containing triadic
closure edges, potentially connecting two nodes if they share a common
neighbor in the substrate network (see Fig.~\ref{fig:diagram}). The
final network is obtained by ``erasing'' the identity of the
edges. i.e. whether they are seminal or due to closure of a
triangle. Conversely, the inference procedure consists in moving in the
opposite direction, i.e. given a simple graph, with no annotations on
the edges, we consider the posterior distribution of all possible
divisions into seminal and triadic closure edges, weighted according to
their plausibility.

We will denote the seminal edges with an adjacency matrix $\A$, and for
its generation we will use the degree-corrected stochastic block model
(DC-SBM)~\cite{karrer_stochastic_2011}, conditioned on a partition $\bb$
of the nodes into $B$ groups, where $b_i\in[1,B]$ is the group
membership of node $i$, which has a marginal distribution given
by~\cite{peixoto_nonparametric_2017}
\begin{multline}\label{eq:sbm}
  P(\A|\bb) = \frac{\prod_{r<s}e_{rs}!\prod_re_{rr}!!\prod_ik_i!}
  {\prod_{i<j}A_{ij}!\prod_iA_{ii}!!\prod_re_r!}\times
  \prod_r\frac{\prod_k\eta_k^r!}{n_r!q(e_r,n_r)}\times\\
  {\frac{B(B+1)}{2} + E - 1 \choose E}^{-1},
\end{multline}
where $e_{rs}=\sum_{ij}A_{ij}\delta_{b_i,r}\delta_{b_j,s}$ is the number
of edges between groups $r$ and $s$ (or twice that for $r=s$),
$e_r=\sum_se_{rs}$, $k_i=\sum_jA_{ij}$ is the degree of node $i$,
$n_r=\sum_i\delta_{b_i,r}$ is the number of nodes in group $r$,
$\eta^r_k=\sum_i\delta_{b_i,r}\delta_{k_i,k}$ is the number of nodes in
group $r$ with degree $k$, $E=\sum_{ij}A_{ij}/2$ is the total number of
edges, and $q(m,n)$ is the number of restricted partitions of integer
$m$ into at most $n$ parts. We refer to
Ref.~\cite{peixoto_nonparametric_2017} for a detailed derivation of this
marginal likelihood, including also the extension for hierarchical
partitions that is straighforward to incorporate, as well as latent
multigraphs~\cite{peixoto_latent_2020} (see
Appendix~\ref{app:multigraph}), both of which we have used in our
analysis. This model is capable of generating networks with arbitrary
degree distributions and mixing patterns between groups of nodes,
including homophily~\cite{peixoto_bayesian_2019}.\footnote{The SBM is
capable of modelling arbitrary kinds of mixing patterns between groups
of nodes, with homophily (or assortative mixing) as a special
case. Therefore our approach is in fact able to disentangle arbitrary
mixing patterns from triadic closure, not only homophily. However,
homophily is the dominant pattern that causes an abundance of triangles,
and hence needs to be distinguished from triadic closure.}

The triadic closure edges are represented by an additional set of $N$
``ego'' graphs $\g$, attributed to each node $u$ of $\A$, where $\g(u)$
is the ego graph of node $u$. The ego graph $\g(u)$ is allowed only to
contain nodes that are neighbors of $u$ in $\A$ (excluding $u$ itself),
and edges that do not exist in $\A$, so that any existing edge in
$\g(u)$ amounts to a triadic closure in $\A$. The adjacency of $\g(u)$
is given by
\begin{equation}
  g_{ij}(u) =
  \begin{cases}
    1, &\text{ if } (i,j) \in \g(u), \\
    0, &\text{ otherwise.}
  \end{cases}
\end{equation}
Let us denote the existence of an open triad $(i,u,j)$ in $\A$ with
\begin{equation}
  m_{ij}(u)=A_{ui}A_{uj}(1-A_{ij}),
\end{equation}
such that $m_{ij}(u) = 1$ if the open triad exists, or $0$ otherwise,
and we adopt the convention $A_{uu} = 0$ throughout. Therefore, an
edge $(i,j)$ can exist in $\g(u)$ only if $m_{ij}(u) = 1$. Based on
this, the ego networks are generated independently with probability,
\begin{equation}\label{eq:p_g}
  P(\g(u)|\A,p_u) = \prod_{i<j}\left[p_um_{ij}(u)\right]^{g_{ij}(u)}\left[1-p_um_{ij}(u)\right]^{1-g_{ij}(u)}.
\end{equation}
where $p_u \in [0,1]$ is a probability associated with node $u$ that
controls the degree to which its neighbors in $\A$ end up connected in
$\g(u)$.  This process may result in the same edge $(i,j)$ existing in
different graphs $\g(u)$, if $i$ and $j$ share more than one common
neighbor in $\A$. We therefore consider the resulting simple graph
$\G(\A,\g)$, constructed by ignoring any multiplicities introduced by
the various ego graphs,
i.e. with adjacency given by
\begin{equation}
  G_{ij}(\A,\g) =
  \begin{cases}
    1, & \text{ if } A_{ij} + \sum_u g_{ij}(u) > 0,\\
    0, & \text{ otherwise.}
  \end{cases}
\end{equation}
The joint probability of the above process is then given by
\begin{equation}
  P(\G,\g,\A|\bm p, \bb) = \bm{1}_{\{\G = \G(\A,\g)\}}P(\A|\bb)\prod_uP(\g(u)|\A,p_u),
\end{equation}
where $\bm{1}_{\{x\}}$ is the indicator function. Unfortunately, the
marginal probability of the final graph
\begin{equation}
  P(\G) = \sum_{\g,\A,\bb}\int P(\G,\g,\A|\bm p, \bb)P(\bm p)P(\bb)\,\dd\bm p,
\end{equation}
with $P(\bm p)$ and $P(\bb)$ being prior probabilities, does not lend
itself to a tractable computation. Luckily, however, this will not be
needed for our inference procedure. Instead, we are interested in the
posterior distribution
\begin{equation}
  P(\g,\A,\bb | \G) = \frac{P(\G,\g,\A|\bb)P(\bb)}{P(\G)},
\end{equation}
which describes the probability of a decomposition of an observed simple
graph $\G$ into its seminal graph $\A$, the underlying community
structure $\bb$, and the triadic closures represented by the ego graphs
$\g$. (Although the marginal distribution $P(\G)$ appears in the
denominator of the above equation, we will see later on that it is
just a normalization constant that does not in fact need to be
computed.) The marginal likelihood
\begin{equation}
P(\G,\g,\A|\bb) = P(\G|\A,\g)P(\g|\A)P(\A|\bb)
\end{equation}
can be computed easily via
\begin{align}
  P(\g|\A) &= \prod_u\int_0^1P(\g(u)|\A,p)P(p)\,\dd p\nonumber\\
  &= \prod_u\left[{\sum_{i<j}m_{ij}(u) \choose \sum_{i<j}g_{ij}(u)}^{-1}
    \frac{1}{1+\sum_{i<j}m_{ij}(u)}\right],\label{eq:marginal_simple}
\end{align}
where we have used a uniform prior $P(p)=1$, omitted for simplicity an
indicator function setting $P(\g|\A)=0$ if $g_{ij} > 0$ and $m_{ij} = 0$
for any $(i,j)$, and with the remaining likelihood term being only the
indicator function, $P(\G|\A,\g)=\bm{1}_{\{\G = \G(\A,\g)\}}$. Although
this choice of priors makes the calculation very simple, it implies that
we expect the observed graphs to always have a large fraction of triadic
closures. In Appendix~\ref{app:general} we describe a slight
modification of this model that makes it more versatile with respect to
the abundance of triadic closures, at the expense of yielding somewhat
longer expressions for the likelihood. We note that we made use of the
modifications specified there in our ensuing analysis, as they can only
improve the use of the model.

\subsection{Iterated triadic closures}

\begin{figure}
  \includegraphics[width=.7\columnwidth]{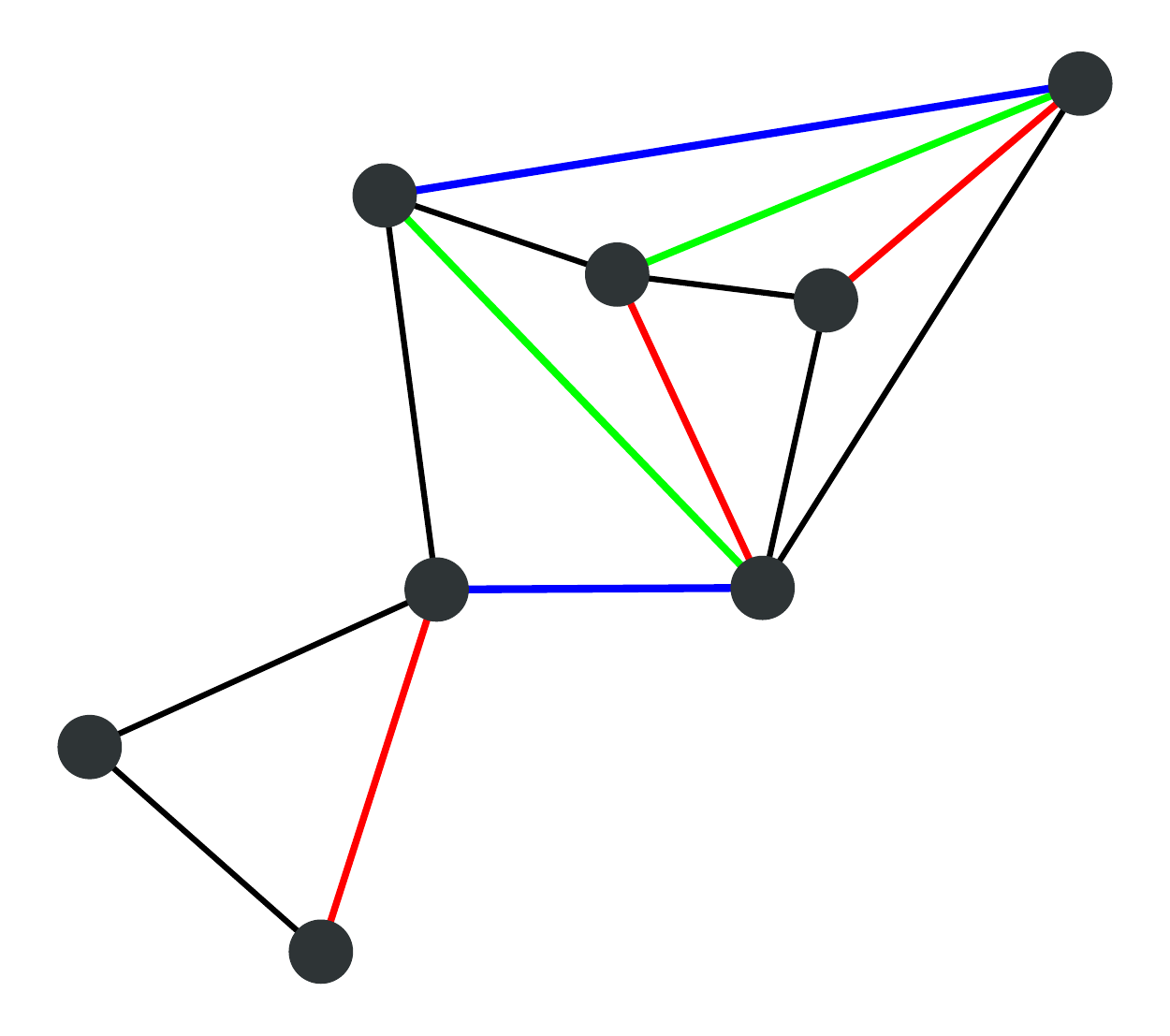} \caption{Example
  network illustrating how iterated triadic closures are implemented in
  the model. The initial network (black edges) receives the first
  generation of triadic closures (red edges). The second generation
  (green edges) can only close triads involving at least one edge of the
  first generation (red). The third generation (blue edges) in turn can
  only close triads involving at least one edge belonging to the second
  generation (green). \label{fig:iteration}}
\end{figure}

Triadic closures increase the number of edges in the network, and in
this way can introduce opportunities for new triadic closures, involving
both older and newer edges. This leads naturally to a dynamical model,
where generations of triadic closures are progressively introduced to
the network.\footnote{One may wonder if such a dynamical process would
also make sense for the homophily part of the model, represented by the
SBM. However, since homophily implies a conditional independence of the
placement of the edges, it does not matter the order with which edges are
added to the network, only their final placement.} We can incorporate this
in our model via ``layers'' of ego graphs $\g^{(l)}$ representing edges
introduced in generation $l\in [1,\dots,L]$. For our formulation, it
will be useful to define the cumulative network at generation $l$,
defined recursively by
\begin{equation}
  A^{(l)}_{ij}=
  \begin{cases}
    1, & \text{ if } A_{ij}^{(l-1)}+\sum_ug_{ij}^{(l)}(u) > 0,\\
    0, & \text{ otherwise,}
  \end{cases}
\end{equation}
with boundary conditions $\A^{(0)}=\A$ (henceforth $\A$ refers solely to
the seminal network, whereas e.g. $\A^{(1)}$ is the resulting network
considering the first iteration of triadic closures, and $\{\A^{(l)}\}$
refers to the set of all generations, including the seminal network),
and $\g^{(0)}(u)$ being empty graphs for all $u$, and we will denote the
final generation as $\A^{(L)}=\G$. The formation of new triadic closure
layers is done according to the probability
\begin{multline}
  P(\g^{(l)}(u)|\A^{(l-1)}, \g^{(l-1)}, p_u^{(l)}) = \\
  \prod_{i<j}\left[p_u^{(l)}m_{ij}^{(l)}(u)\right]^{g_{ij}^{(l)}(u)}
  \left[1-p_u^{(l)}m_{ij}^{(l)}(u)\right]^{1-g_{ij}^{(l)}(u)}.
\end{multline}
where an open triad $(i,u,j)$ at generation $l$ is denoted by
\begin{equation}\label{eq:ml}
  m_{ij}^{(l)}(u) = w_{ij}^{(l)}(u)\left(1-A_{ij}^{(l-1)}\right),
\end{equation}
so that $m_{ij}(u) \in \{0, 1\}$, where
\begin{multline}\label{eq:wl}
  w_{ij}^{(l)}(u) =\\
  \begin{cases}
    1, & \text{ if } A_{ui}^{(l-1)}\sum_vg^{(l-1)}_{uj}(v) + A_{uj}^{(l-1)}\sum_vg^{(l-1)}_{ui}(v) > 0,\\
    0, & \text{ otherwise,}
  \end{cases}
\end{multline}
determines whether or not the open triad $(i,u,j)$ at generation $l$ has
at least one of the edges $(u,i)$ or $(u,j)$ formed exactly at the
preceding generation $l-1$. This restriction means that triadic closures
at generation $l$ can only close new triads that have been introduced at
generation $l-1$, not previously. The reason for this is a matter of
identifiability: an edge at generation $l$ that closes an open triad
that has been formed at generation $l' < l$ could also have been
generated in any of the intermediate generations $[l',l-1]$, thus
introducing an inevitable ambiguity in the inference. The above
restriction removes the ambiguity, and forces the new generations to
form triadic closures which could not have existed in the preceding
generations (see Fig.~\ref{fig:iteration}). Note that this restriction
does not significantly alter the generality of the model, since the same
final networks can still be formed with the similar probability despite
it.\footnote{In more detail, we can recover the unconstrained model by
substituting Eq.~\ref{eq:ml} with $m_{ij}^{(l)}(u) =
A_{ui}^{(l-1)}A_{uj}^{(l-1)}(1-A_{ij}^{(l-1)})$. Since the edges at each
layer $l$ are generated independently, this would only mean that more
edges would be generated on top of each other across the layers. Since
these multiple edges are removed in the end, this means that the
unconstrained model would have a higher probability for forming edges
that are possible in the earlier generations, since they could appear
also in the later ones. But since we typically require only a very small
number of generations, this is a very minor effect, and both models
become very similar, while the constrained model is easier to infer.}

With the above, the joint likelihood of all generations is given by
\begin{multline}
  P(\{\g^{(l)}\},\{\A^{(l)}\}|\bb,\bm p) =\\
  P(\A|\bb)\prod_{l=1}^{L}\prod_uP(\g^{(l)}(u)|\A^{(l-1)}, \g^{(l-1)}, p_u^{(l)}).
\end{multline}
Following the same calculation as before, we obtain the individual
marginal likelihood at each generation $l$ as
\begin{equation}
  P(\{\g^{(l)}\},\{\A^{(l)}\}|\bb) =
  P(\A|\bb)\prod_{l=1}^{L}P(\g^{(l)}|\A^{(l-1)}, \g^{(l-1)}).
\end{equation}
with the individual terms in the product being entirely analogous to
Eq.~\ref{eq:marginal_simple},
\begin{multline}
  P(\g^{(l)}|\A^{(l-1)}, \g^{(l-1)}) = \\
  \prod_u\left[{\sum_{i<j}m_{ij}^{(l)}(u) \choose \sum_{i<j}g_{ij}^{(l)}(u)}^{-1}
    \frac{1}{1+\sum_{i<j}m_{ij}^{(l)}(u)}\right].
\end{multline}
Finally, the posterior distribution for the reconstruction becomes
\begin{equation}\label{eq:posterior}
  P(\{\g^{(l)}\},\{\A^{(l)}\},\bb | \G) = \frac{P(\G,\{\g^{(l)}\},\{\A^{(l)}\}|\bb)P(\bb)}{P(\G)}.
\end{equation}
Note that for $L=1$ we recover the previous model. Having to specify $L$
beforehand is not a strict necessity, since the inference will only
occupy new generations if this yields a more parsimonious description of
the network.~\footnote{If we wanted to tread $L$ as an unknown, we
should introduce a prior for $L$, $P(L)$, and include that in the
posterior as well. However, with the parametrization in
Appendix~\ref{app:general}, generations which are unpopulated with edges
have no contribution to the marginal likelihood. Therefore we can
simply set $L$ to be a sufficiently large value, for example
$L={N\choose 2}$, since for later generations it is impossible to add
new edges.}

\subsection{Inference algorithm}
\begin{figure*}
  \begin{tabular}{P{.33\textwidth}P{.33\textwidth}P{.33\textwidth}}
    \includegraphics[width=.33\textwidth]{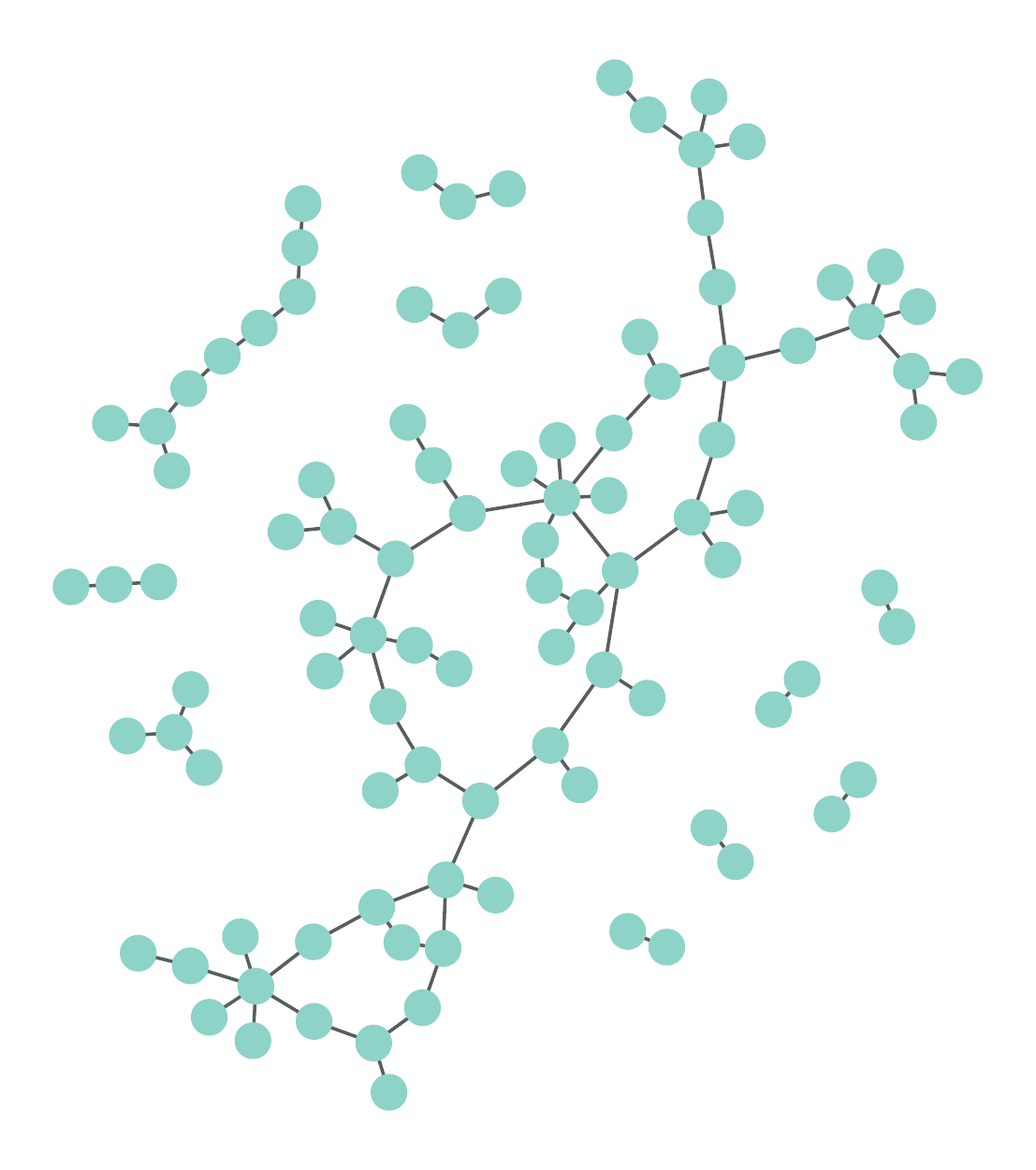} &
    \includegraphics[width=.33\textwidth]{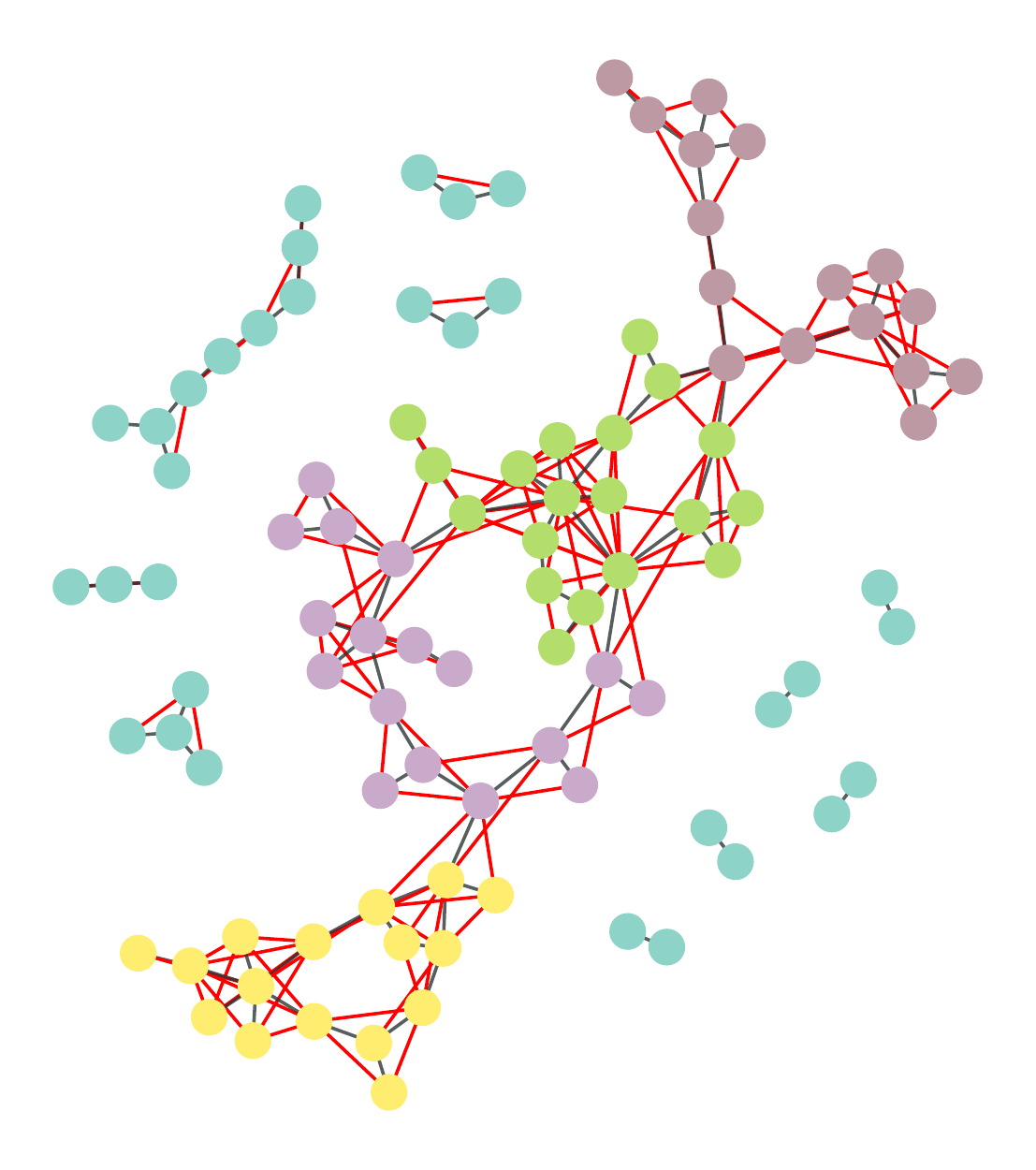}&
    \includegraphics[width=.33\textwidth]{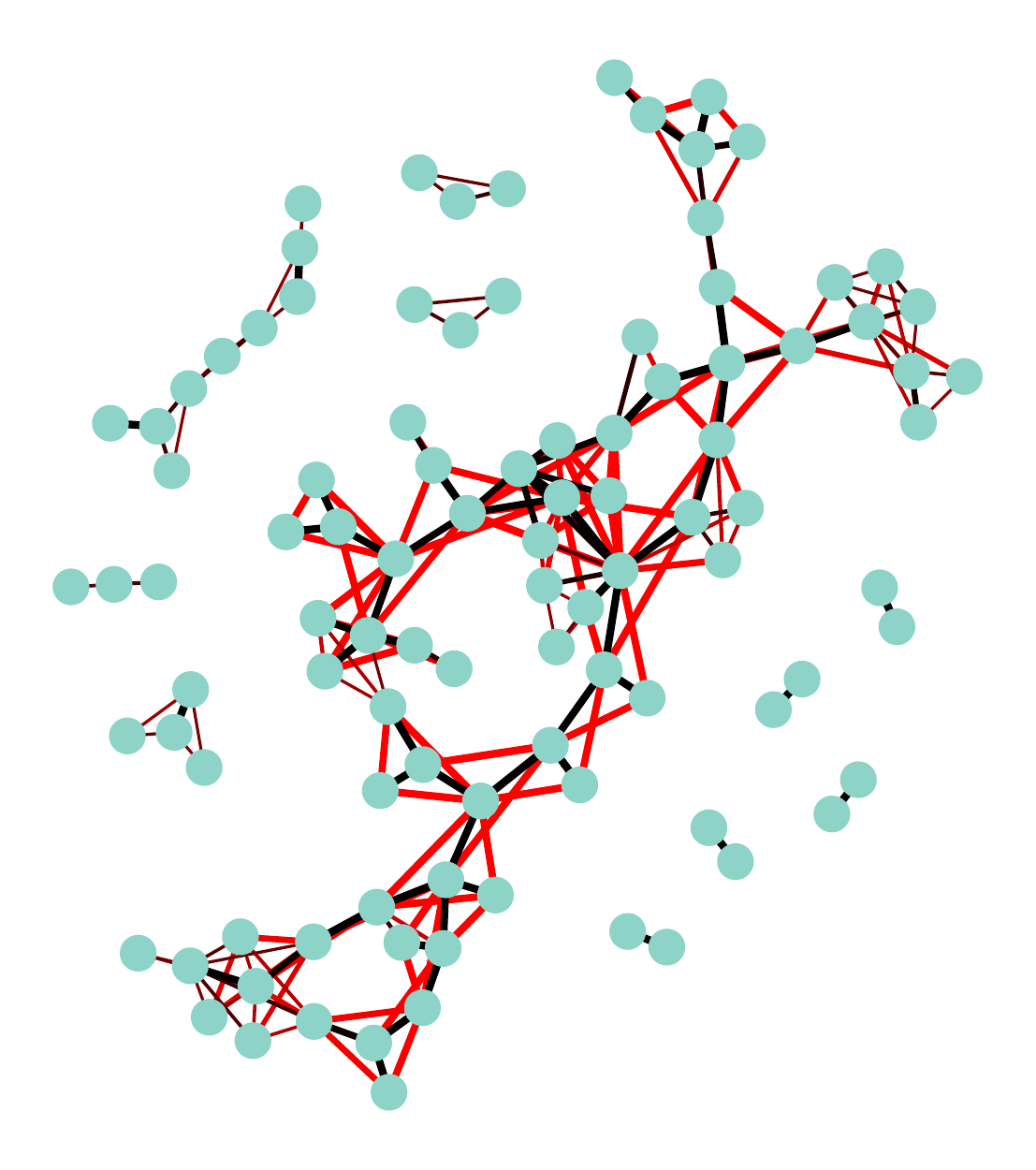}\\
    (a) Random seminal edges&
    (b) Triadic closure edges and spurious communities found with SBM ($\Sigma_{\text{SBM}}=801.7$ nats)&
    (c) Inference of the SBM/TC model ($\Sigma_{\text{SBM/TC}}=590.6$ nats)
  \end{tabular} \caption{\textbf{(a)} Example artificial network
  generated as a fully random graph with a geometric degree
  distribution, $N=100$ nodes and $E=94$ edges, and \textbf{(b)} a
  process of triadic closure based on network (a) with parameter
  $p_u=0.8$ for every node, with closure edges shown in red. It is also
  shown the partition found by fitting the SBM to the resulting network,
  and the description length obtained. \textbf{(c)} The result of
  inferring the SBM/TC model, which uncovers a single partition --- no
  community structure --- and the closure edges shown in red (the
  thickness of the edges correspond to the marginal probabilities
  $\pi_{ij}$ and $1-\pi_{ij}$ for the seminal and closure edges,
  respectively). It is also shown the description length of the SBM/TC
  fit. \label{fig:random}}
\end{figure*}

The posterior distribution of Eq.~\ref{eq:posterior} can be written
exactly, up to a normalization constant. However, this fact alone does
not allow us to directly sample from this distribution, which can only
be done in very special cases. Instead, we rely here on Markov chain
Monte Carlo (MCMC), implemented as follows. We begin with an arbitrary
choice of $\{\g^{(l)}\}$, $\{\A^{(l)}\}$ and $\bb$ that is compatible with our
observed graph $\G$. We then consider modifications of these quantities,
and accept or reject them according to the Metropolis-Hastings
criterion~\cite{metropolis_equation_1953,hastings_monte_1970}. More
specifically, we consider moves of the kind
$P(\{{\g'}^{(l)}\},\{{\A^{(l)}}'\}|\{{\g}^{(l)}\},\{\A^{(l)}\})$, and accept them
according to the probability
\begin{multline}
\min\left(1,
\frac{P(\{{\g'}^{(l)}\},\{{\A^{(l)}}'\},\bb | \G)}
     {P(\{\g^{(l)}\},\{\A^{(l)}\},\bb | \G)}\times\right.\\
     \left.\frac{P(\{{\g}^{(l)}\},\{\A^{(l)}\}|\{{\g'}^{(l)}\},\{{\A^{(l)}}'\})}{P(\{{\g'}^{(l)}\},\{{\A^{(l)}}'\}|\{{\g}^{(l)}\},\{\A^{(l)}\})}\right)\\
\end{multline}
which, as we mentioned before, does not require the computation of the
intractable marginal probability $P(\G)$. We also consider moves that
change the community structure, according to a proposal
$P(\bb'|\bb)$ and accept with probability
\begin{equation}
  \min\left(1,
  \frac{P(\A|\bb')P(\bb')P(\bb|\bb')}
       {P(\A|\bb)P(\bb)P(\bb'|\bb)}\right).
\end{equation}
For the latter we use the merge-split moves described in
Ref.~\cite{peixoto_merge-split_2020}. Iterating the moves described
above eventually produces samples from the target posterior
distribution. In Appendix~\ref{app:mcmc} we specify the details of the
particular move proposals we use.

Given samples from the posterior distribution, we can use them to
summarize it in a variety of ways. A useful quantity is the marginal
probability $\pi_{ij}$ of an edge $(i,j)$ being seminal, which is given
by
\begin{equation}
  \pi_{ij} = \sum_{\{\g^{(l)}\},\{\A^{(l)}\},\bb}A_{ij}P(\{\g^{(l)}\},\{\A^{(l)}\},\bb | \G).
\end{equation}
Conversely, the reciprocal quantity,
\begin{equation}
  1-\pi_{ij},
\end{equation}
corresponds to the probability that edge $(i,j)$ is due to triadic
closure, occurring in any generation or ego graph. Therefore, the
quantity $\bm\pi$ gives us a concise summary of posterior decomposition
of a network, and we will use it throughout our analysis. (It is easy to
devise and compute other summaries, such as the marginal probability of
an edge belonging to a given triadic generation, or a particular ego
graph, but we will not have use for those in our analysis.)

\section{Distinguishing community structure from triadic closure}\label{sec:artificial}

Here we illustrate how triadic closure can be mistaken as community
structure, and how our inference method is capable of uncovering it. We
begin by considering an artificial example, where we first sample a
fully random network with a geometric degree distribution, $N=100$ nodes
and $E=94$ edges, as shown in Fig.~\ref{fig:random}a. This network does
not possess any community structure, since the probability of observing
an edge is just proportional to the product of the degrees of the
endpoint nodes --- indeed if we fit a DC-SBM to it, we uncover,
correctly, only a single group. Conditioned on this network,
Fig.~\ref{fig:random}b shows sampled triadic closure edges, according to
the model described previously, where each node has the same probability
$p_u=0.8$ of having neighbors connected in their ego graphs. In the
same figure we show the result of fitting the DC-SBM on the network
obtained by ignoring the edge types. That approach finds five
assortative communities, corresponding to regions of higher densities of
edges induced by the random introduction of transitive edges. One should
not, however, interpret the presence of these regions as a special
affinity between the respective groups of nodes, since they are a result
of a random process that has no relation to that particular division of
the network --- indeed, if we run the whole process again from the
beginning, the nodes will most likely end up clustered in completely
different ``communities.'' If we now perform the inference of our SBM
with triadic closure (SBM/TC), we obtain the result shown in
Fig.~\ref{fig:random}c. Not only are we capable of distinguishing the
seminal from the triadic closure edges (AUC ROC = $0.92$), but we also
correctly identify the presence of a single group of nodes, which is in
full accordance with the completely random nature in which the network
has been generated. In other words, with the SBM/TC we are not misled by
the density heterogeneity introduced by triadic closures into thinking
that the network possesses real community structure, and we realize
instead that they can be better explained by a different process.

In the artificial example considered above, the result obtained with the
SBM/TC model is more appealing, since it more closely matches the known
generative process that was used. However, in more realistic situations,
we will need to decide if it provides a better description of the data
without such privileged information. In view of this, we can make our
model selection argument more formal in the following way. Suppose we
are considering a partition $\bb^{(1)}$ found with inferring the SBM on
a given network, as well as another partition $\bb^{(2)}$ and ego graphs
$\{\g^{(l)}\}$ found with the SBM/TC model.  We can decide which one
provides a better description of a network via the posterior odds ratio,
\begin{align}
  \Lambda
  &= \frac{P(\bb^{(2)}, \{\g^{(l)}\}, \mathcal{H}_{\text{SBM/TC}}|\G)}
      {P(\bb^{(1)}, \mathcal{H}_{\text{SBM}}|\G)}\\
  &= \frac{P(\G,\{\g^{(l)}\},\{\A^{(l)}\},\bb^{(2)})}
      {P(\G,\bb^{(1)})}
      \times
      \frac{P(\mathcal{H}_{\text{SBM/TC}})}{P(\mathcal{H}_{\text{SBM}})},
\end{align}
where $P(\mathcal{H}_{\text{SBM/TC}})$ and $P(\mathcal{H}_{\text{SBM}})$
are the prior probabilities for either model. In case these are the
same, we have
\begin{equation}
  \Lambda = \mathrm{e}^{-(\Sigma_{\text{SBM/TC}} - \Sigma_{\text{SBM}})},
\end{equation}
where $\Sigma_{\text{SBM/TC}}$ and $\Sigma_{\text{SBM}}$ are the description
lengths of both hypotheses, given by
\begin{align}
  \Sigma_{\text{SBM/TC}} &= -\ln P(\G,\{\g^{(l)}\},\{\A^{(l)}\},\bb^{(2)}),\\
  \Sigma_{\text{SBM}} &= -\ln P(\G,\bb^{(1)}).
\end{align}
The description length~\cite{grunwald_minimum_2007} measures the amount
of information necessary to encode both the data and the model
parameters, and hence accounts both for the quality of fit and the model
complexity. The above means that the model that is most likely \emph{a
posteriori} is the one that \emph{most compresses} the data under its
parametrization, and thus the criterion amounts to an implementation of
Occam's razor, since it points to the best balance between model
complexity and fitness.

Before we employ the above criterion to select between both models
considered, it is important to emphasize that the pure SBM is ``nested''
inside the SBM/TC, since the former amounts to the special case of the
latter when there are zero triadic closure edges. In particular, if we
use the more general parametrization described in
Appendix~\ref{app:general}, in the situation with zero triadic edges,
i.e. all $\{\g^{(l)}\}$ are empty graphs $\g_{\text{empty}}$ and
$\A=\G$, we have
\begin{equation}
  P(\G,\{\g^{(l)}=\g_{\text{empty}}\},\A=\G,\bb) \geq \frac{P(\G,\bb)}{N+1}.
\end{equation}
Therefore, in general, we must have
\begin{multline}
  \underset{\{\g^{(l)}\},\{\A^{(l)}\},\bb}{\max}\; \ln P(\G,\{\g^{(l)}\},\{\A^{(l)}\},\bb) \geq\\
  \underset{\bb}{\max}\; \ln P(\G,\bb) - \ln(N+1).
\end{multline}
Since the last logarithm term becomes negligible for large networks,
typically the use of the SBM/TC can only reduce the description length
of the data. Therefore, in situations where there is no evidence for
triadic closure, both models should yield approximately the same
description length value.

In Fig.~\ref{fig:random} we show the description lengths for both models
for the particular example discussed previously, where we can see that
the SBM/TC provides a substantially better compression of the data,
therefore yielding a more parsimonious and hence more probable account
of the how the data was generated --- which happens also to be the true
one in this controlled setting.

\begin{figure}
  \begin{tabular}{cc}
    \multicolumn{2}{c}{Inference using SBM}\\
    \begin{overpic}[width=.5\columnwidth]{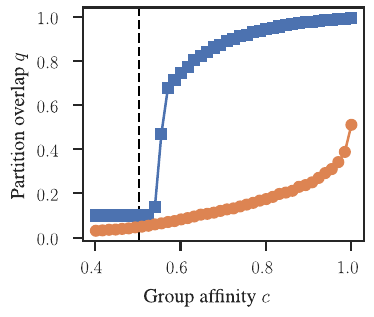}
      \put(0,0){(a)}
    \end{overpic}&
    \begin{overpic}[width=.5\columnwidth]{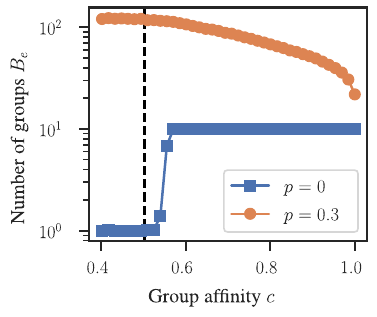}
      \put(0,0){(b)}
    \end{overpic}\\[1em]
    \multicolumn{2}{c}{Inference using SBM/TC}\\
    \begin{overpic}[width=.5\columnwidth]{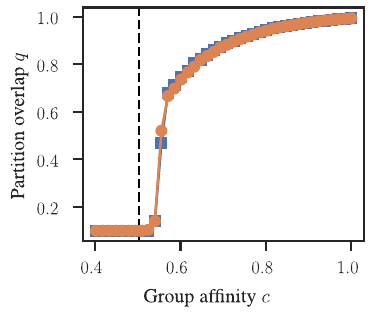}
      \put(0,0){(c)}
    \end{overpic}&
    \begin{overpic}[width=.5\columnwidth]{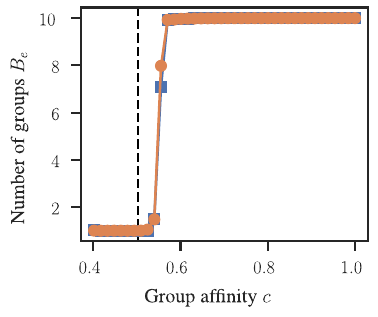}
      \put(0,0){(d)}
    \end{overpic}
  \end{tabular} \caption{\label{fig:recovery}Recovery of community
  structure for artificial networks generated from the PP model with
  added triadic closure, as described in the text, for networks with
  $N=10^4$ nodes, average degree $\avg{k}=5$, $B=10$ planted groups, and
  uniform triadic closure probability $p_u=p$ shown in the
  legend. Figures (a) and (b) correspond to inferences done using the
  SBM, and (c) and (d) with the SBM/TC model. All results where averaged
  over 10 network realizations. The vertical dashed line marks the
  detectability transition value $c^*_+$, described in the text.}
\end{figure}

We proceed with a more systematic analysis of how triadic closure can
interfere in community detection with artificial networks generated by
the SBM, more specifically the special case known as the planted
partition model (PP), where the $B$ groups have equal size, and the
number of edges between groups is given by
\begin{equation}
  e_{rs} = 2E\left[\frac{c}{B}\delta_{rs} + \frac{1-c}{B(B-1)}(1-\delta_{rs})\right],
\end{equation}
where $c\in[0,1]$ determines the affinity between the (dis)assortative
groups. For this model, we know that there are critical values
\begin{equation}
  c^*_{\pm} = \frac{1}{B} \pm \frac{B-1}{B\sqrt{\avg{k}}},
\end{equation}
such that if $c\in [c^*_{-}, c^*_{+}]$ then no algorithm can infer a
partition that is correlated to the true one from a single network
realization, as it becomes infinitely large
$N\to\infty$~\cite{decelle_asymptotic_2011}. Starting from a network
generated with the PP model, we include triadic closure edges via the
global probability $p_u=p$ for every node in the network. Based on the
resulting network, we attempt to recover the original communities, using
the SBM and the SBM/TC model. A result of this analysis is shown in
Fig.~\ref{fig:recovery}, where we compute the maximum
overlap~\cite{peixoto_revealing_2021} $q\in[0,1]$ between the inferred
$\hat\bb$ and true partition $\bb$, defined as
\begin{equation}
  q = \underset{\mu}{\max}\;\frac{1}{N}\sum_i\delta_{\mu(\hat b_i), b_i},
\end{equation}
where $\mu(r)$ is a bijection between the group labels in $\hat\bb$ and
$\bb$, as well as the effective number of inferred groups
$B_e=\mathrm{e}^S$, where $S$ is the group label entropy
\begin{equation}\label{eq:eB}
  S = -\sum_r \frac{n_r}{N}\ln \frac{n_r}{N}.
\end{equation}
As can be seen in Fig.~\ref{fig:recovery}a, the presence of triadic
closure edges can have a severe negative effect on the recovery of the
original partitions when using the SBM. In Fig.~\ref{fig:recovery}b we
see that the number of groups uncovered can be orders of magnitude
larger than the original partition, specially when the latter is not
even detectable. This shows that the apparent communities that arise out
of the formation of triangles substantially overshadow the underlying
true community structure. The situation changes considerably when we use
the SBM/TC instead, as shown Fig.~\ref{fig:recovery}c. In this case, the
presence of triadic closure has no noticeable effect on the
detectability of the true community structure, and we obtain a recovery
performance indistinguishable from the SBM in the case with no
additional edges. As seen in Fig.~\ref{fig:recovery}c the same is true
for the number of groups inferred. These results seem to point to a
robust capacity of the SBM/TC model to reliably distinguish between
actual community structure, and the density fluctuations with result
from triadic closures.

\section{Empirical networks}\label{sec:empirical}
\begin{figure}
  \begin{tabular}{P{.5\columnwidth}P{.5\columnwidth}}
    \includegraphics[width=.5\columnwidth]{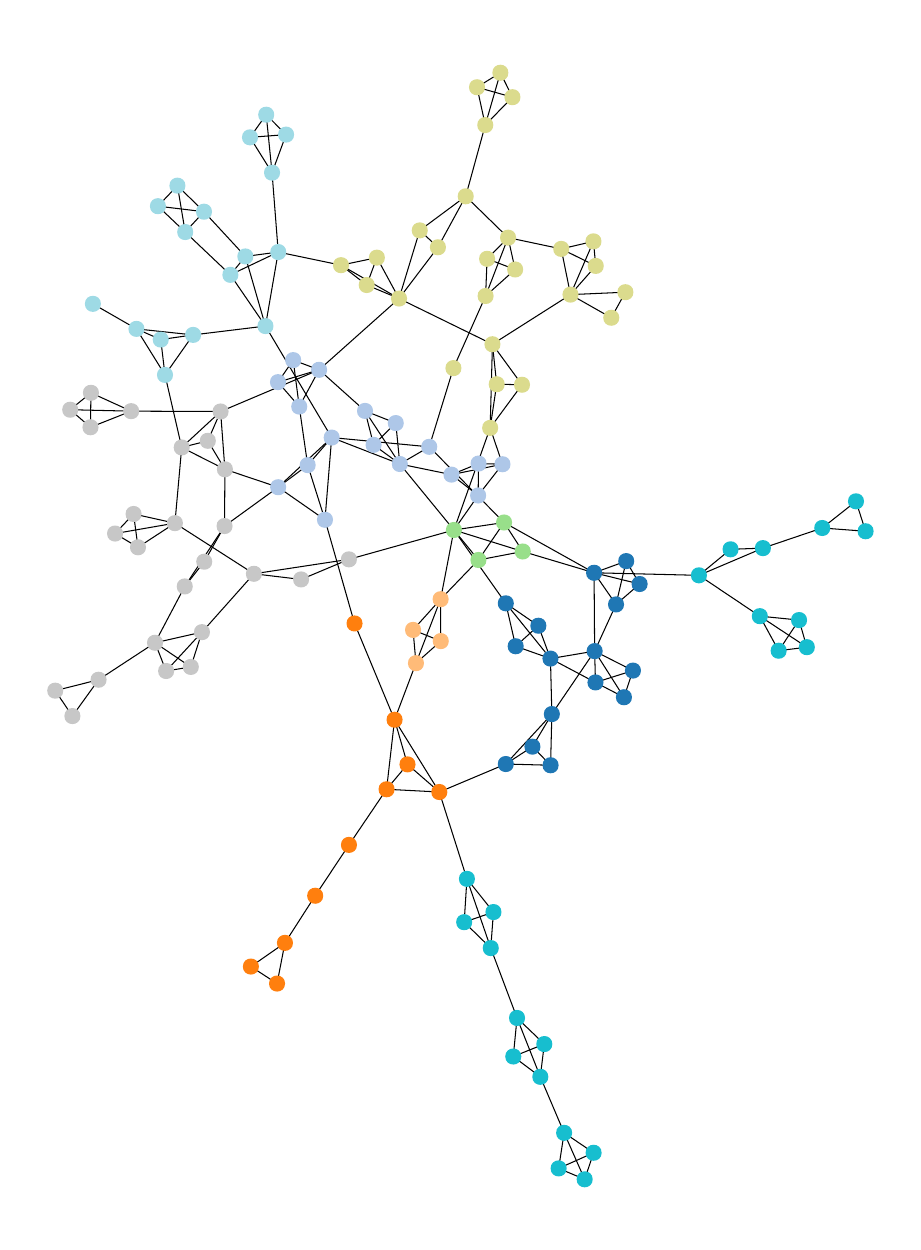} &
    \includegraphics[width=.5\columnwidth]{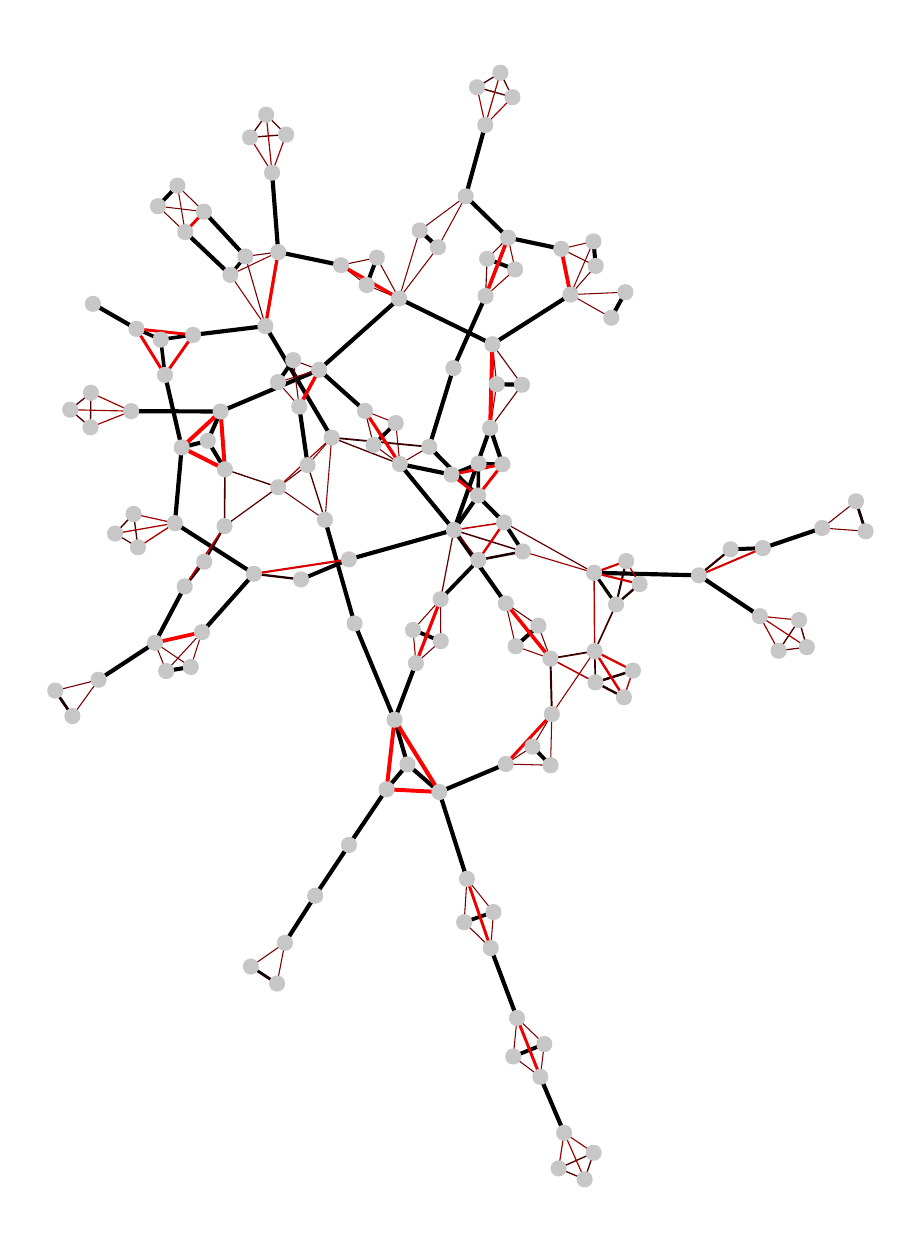}\\
    (a) SBM, $\Sigma_{\text{SBM}}=1145.6$ nats & (b) SBM/TC, $\Sigma_{\text{SBM/TC}}=935.1$ nats
  \end{tabular}

  \caption{Network of cooperation between students~\cite{Fire_2012}. (a)
  Fit of the SBM, yielding $B=9$ communities. (b) Fit of the SBM/TC,
  uncovering a single community, and triadic closure edges shown in
  red. The thickness of the edges correspond to the marginal
  probabilities $\pi_{ij}$ and $1-\pi_{ij}$ for the seminal and closure
  edges, respectively.\label{fig:student_cooperation}}
\end{figure}

\begin{figure*}
  \centering
  {\larger Friendships between high school students}
  \begin{tabular}{P{.33\textwidth}P{.33\textwidth}P{.33\textwidth}}
    \begin{overpic}[width=.33\textwidth]{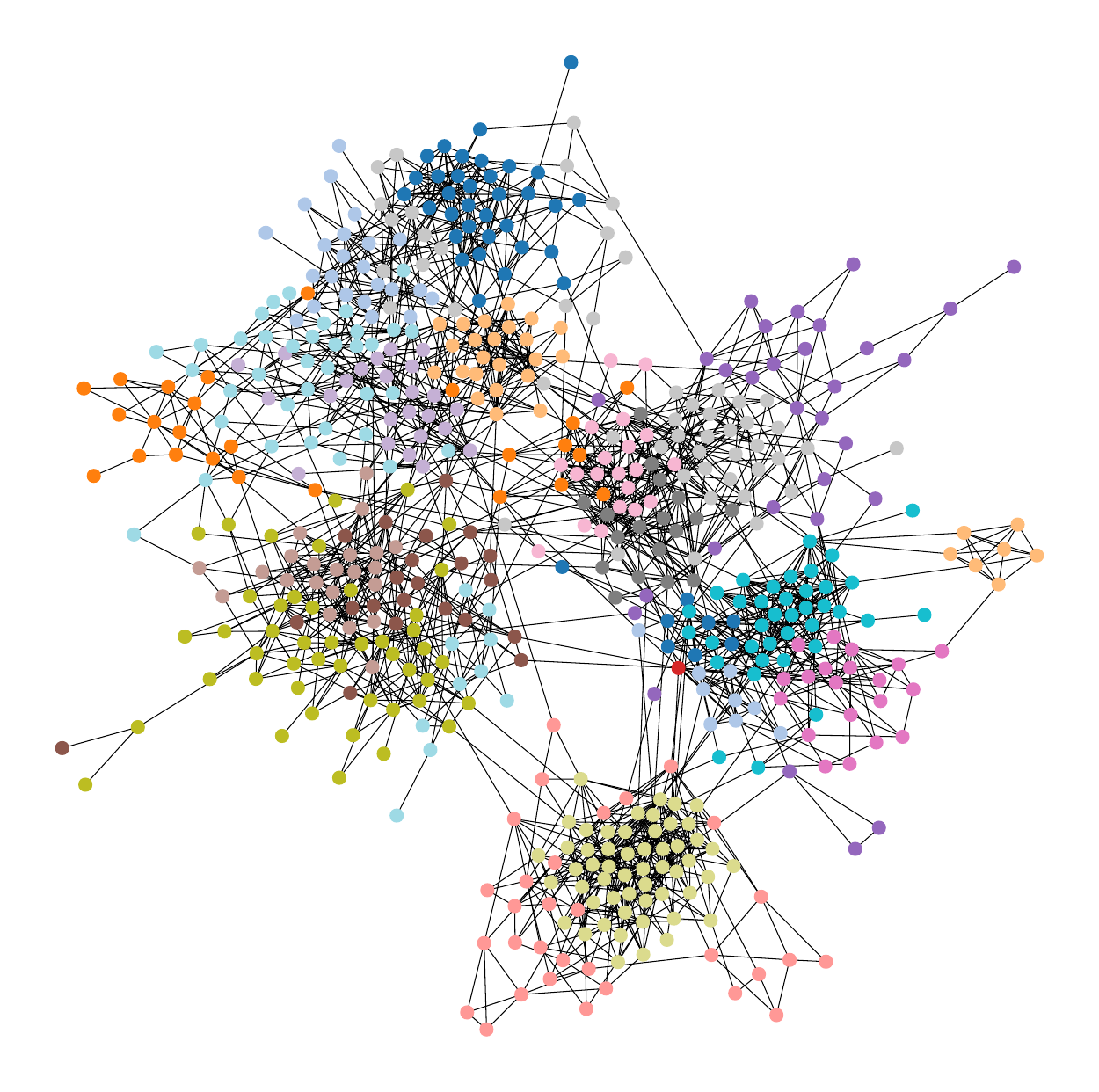}
      \put(95,55){\color{black}$\bm{\swarrow}$}
    \end{overpic} &
    \begin{overpic}[width=.33\textwidth]{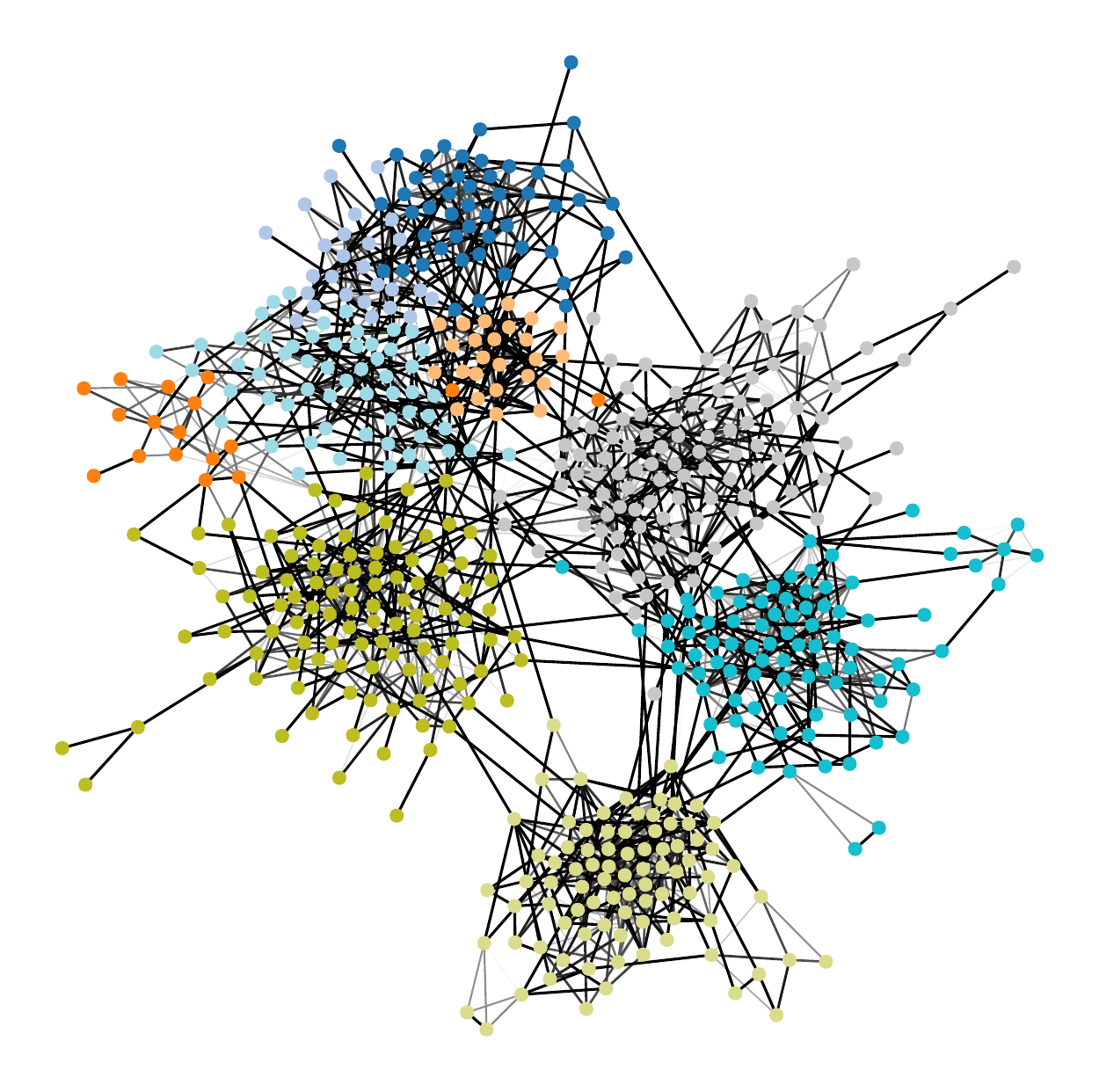}
      \put(0,0){\includegraphics[width=.33\textwidth]{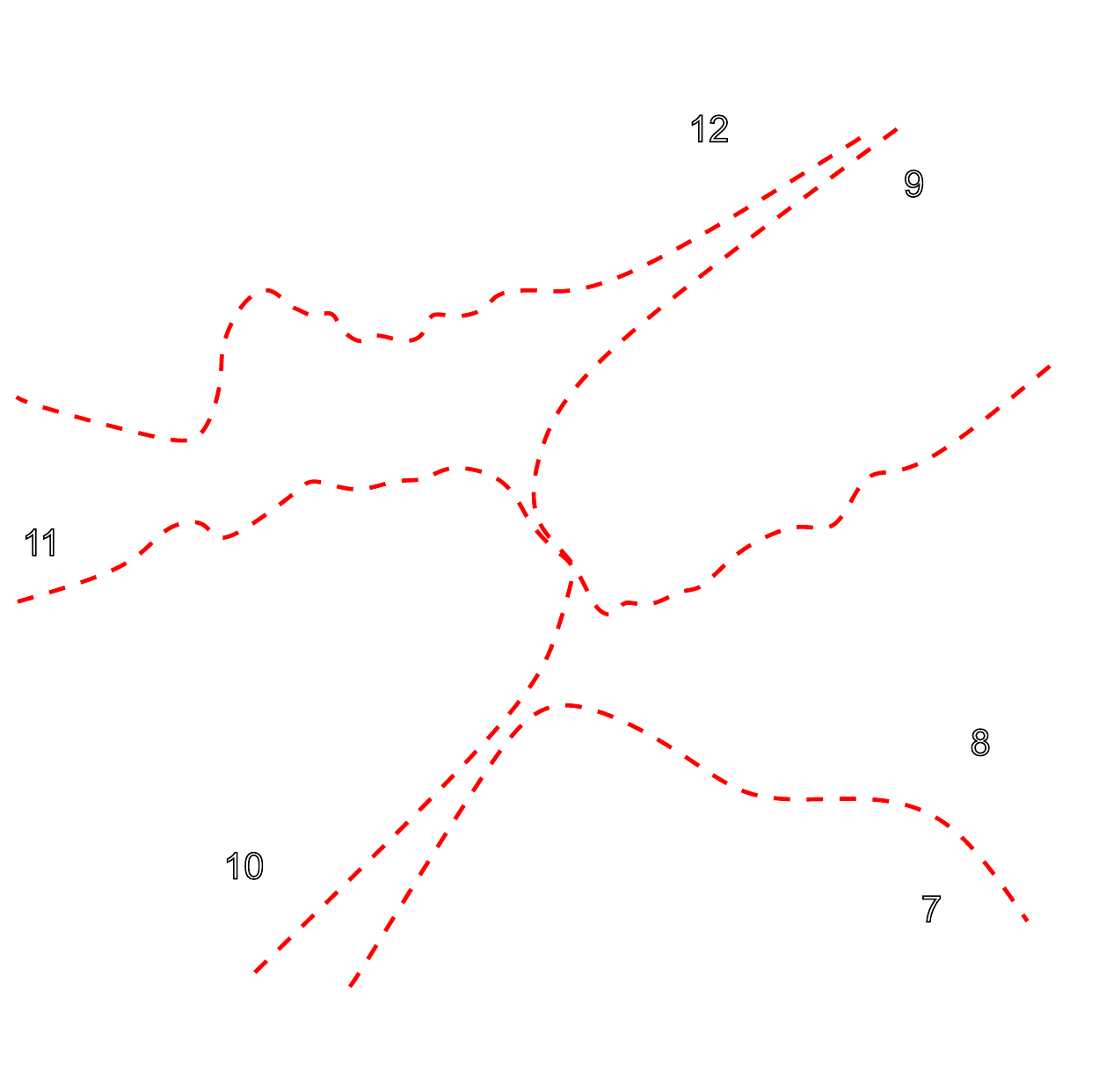}}
    \end{overpic}&
    \includegraphics[width=.33\textwidth]{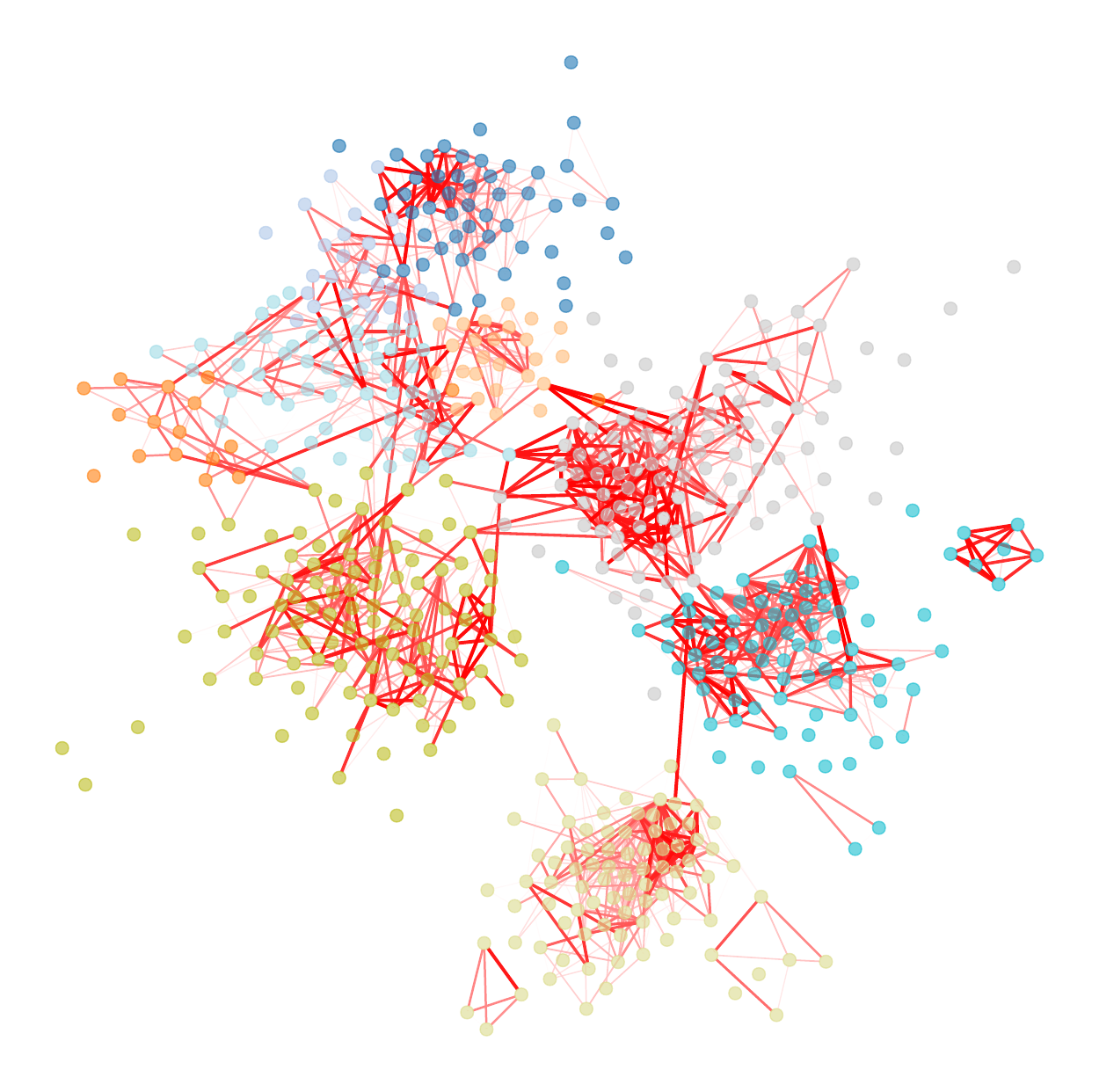}\\
    & Seminal edges & Triadic closure edges \\[.3em]
    (a) SBM ($\Sigma_{\text{SBM}}=8757.0$ nats) & \multicolumn{2}{c}{ (b) SBM/TC ($\Sigma_{\text{SBM/TC}}=8456.3$ nats)}
  \end{tabular}

  \vspace{.5em}
  {\larger Collaborations between network scientists}
  \begin{tabular}{cc}
    \includegraphics[width=.5\textwidth]{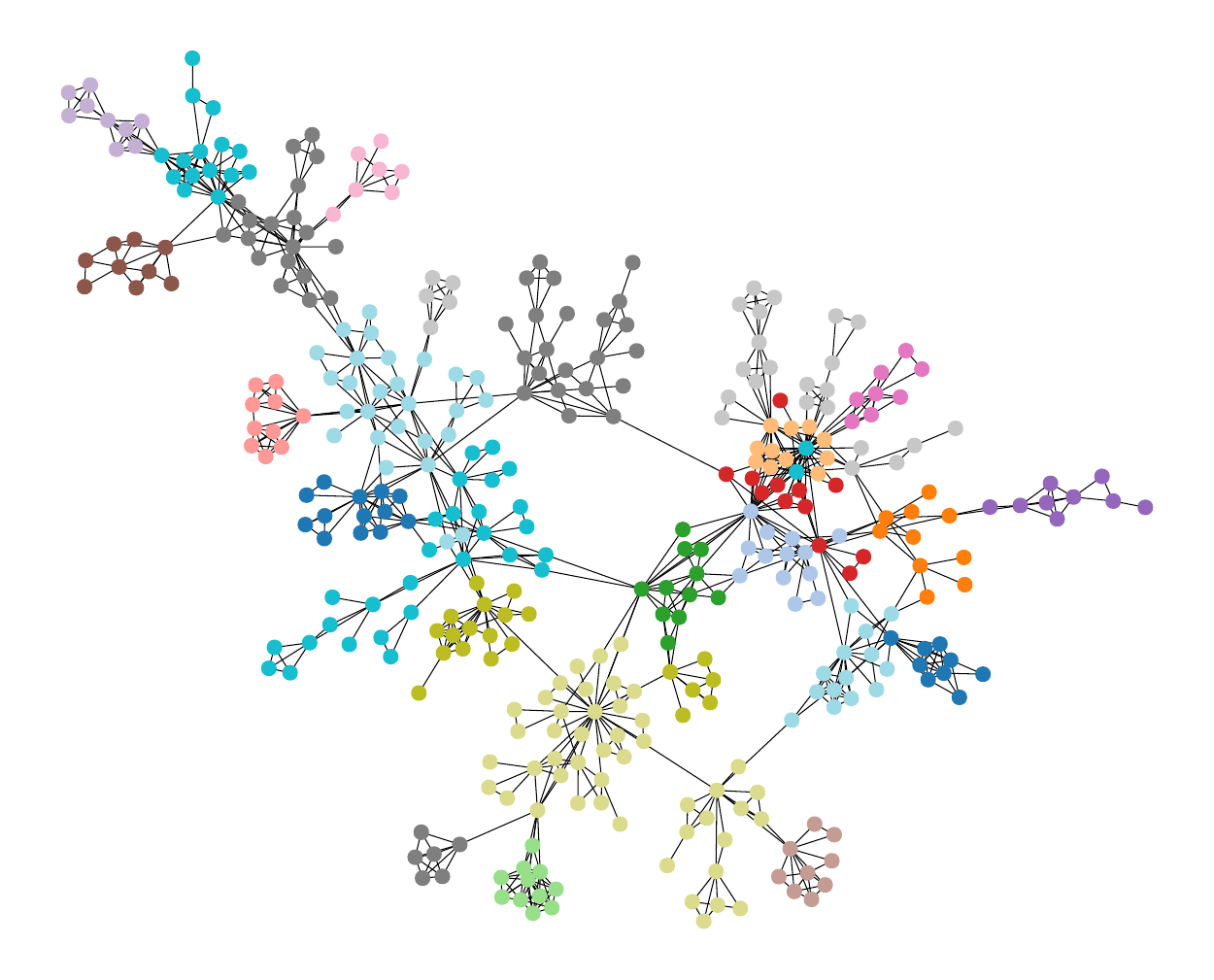} &
    \includegraphics[width=.5\textwidth]{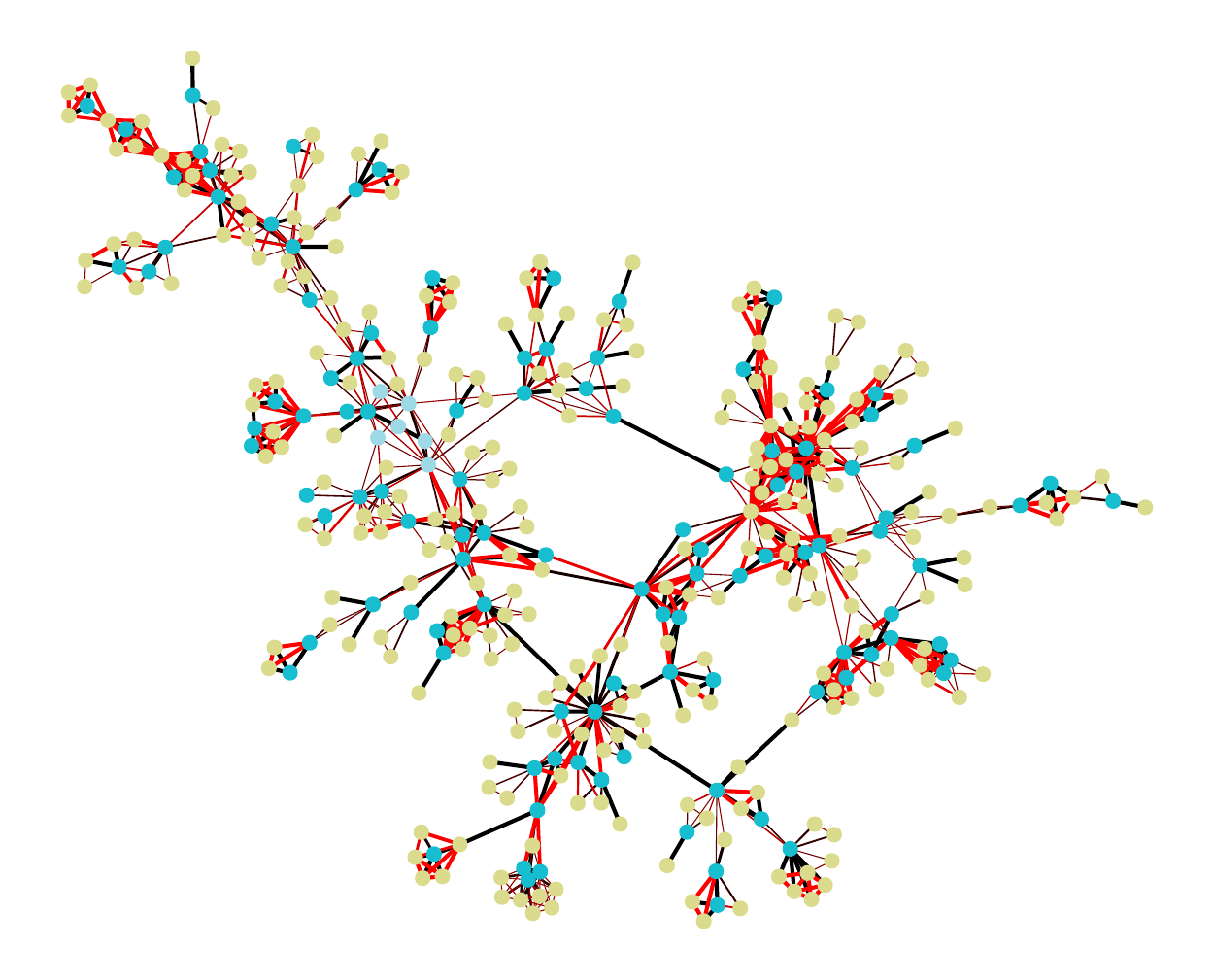}\\
    (c) SBM, $\Sigma_{\text{SBM}}=3816.3$ nats &
    (d) SBM/TC, $\Sigma_{\text{SBM/TC}}=3009.9$ nats
  \end{tabular} \caption{(Top panel) Network of friendships between high school students ---
  Adolescent health (comm26)~\cite{Moody_2001}.  (a) Fit of the SBM,
  yielding $B=26$ communities. (b) Fit of the SBM/TC, uncovering $B=9$
  communities, with seminal (black) and triadic closure (red) edges
  shown separately in the left and right panels.
  (Bottom Panel) Network of collaborations between network
  scientists~\cite{Newman_2006}. (c) Fit of the SBM, yielding $B=27$
  communities. (d) Fit of the SBM/TC, uncovering only $B=3$ groups, and
  triadic closure edges shown in red. The thickness of the edges
  correspond to the marginal probabilities $\pi_{ij}$ and $1-\pi_{ij}$
  for the seminal and closure edges,
  respectively.\label{fig:netsci_add_health}}
\end{figure*}

\begin{figure}
  \begin{tabular}{P{.5\columnwidth}P{.5\columnwidth}}
    \includegraphics[width=.5\columnwidth]{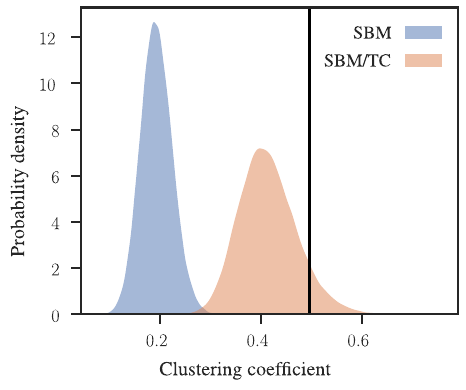} &
    \includegraphics[width=.5\columnwidth]{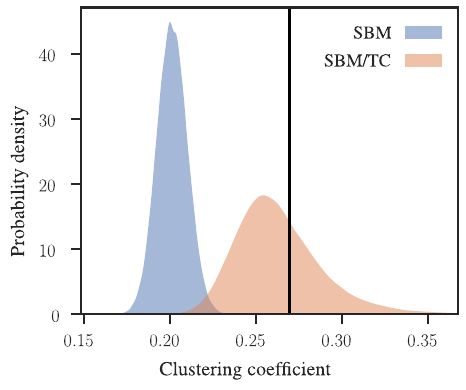}\\
    {\smaller (a) Cooperation between students } & {\smaller (b) Adolescent health (comm26)}\\
    \includegraphics[width=.5\columnwidth]{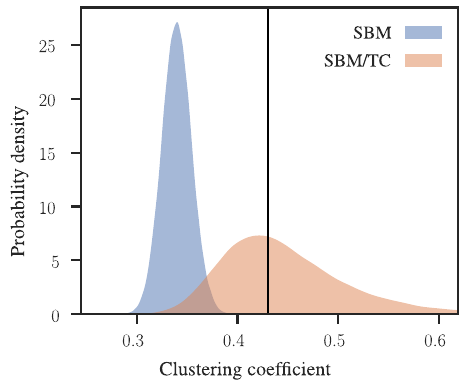} &
    \includegraphics[width=.5\columnwidth]{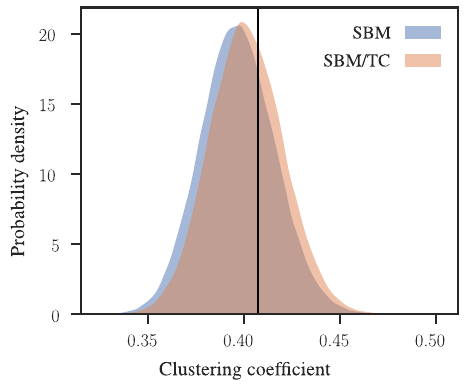}\\
    {\smaller (c) Scientific collaborations in Network Science } & {\smaller(d) NCAA college football 2000}
  \end{tabular}

  \caption{Posterior predictive distributions of the clustering
  coefficient, as described in the text, for the SBM and SBM/TC as
  indicated in the legend, for different datasets. The vertical line
  shows the empirical value $C(\G)$.\label{fig:c-pred}}
\end{figure}

\begin{figure}
  \includegraphics[width=.4\textwidth]{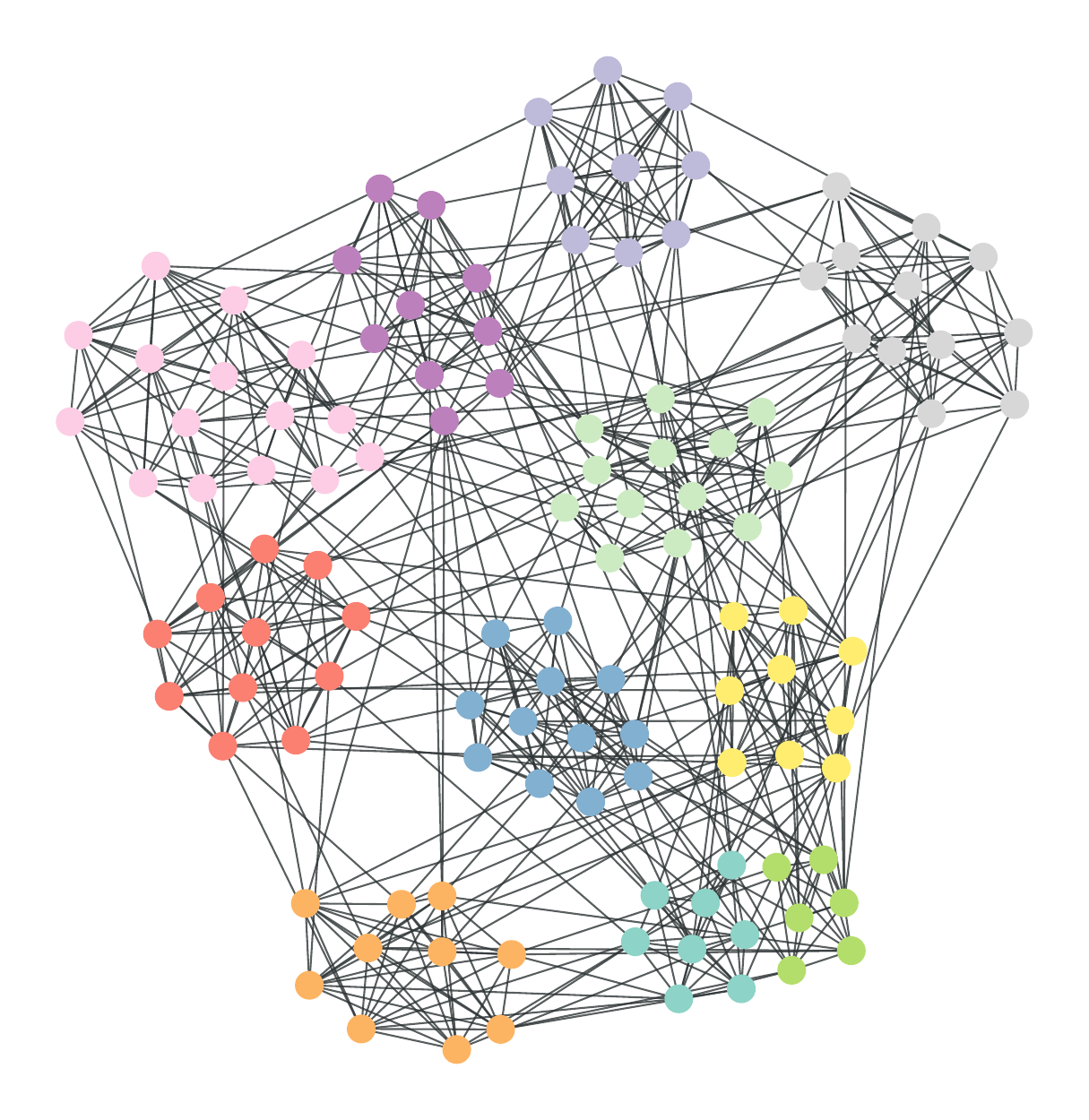}

  \caption{Network of games between American college football
    teams (NCAA college football 2000)~\cite{Girvan_2002}. The node
    colors show the fit of the SBM and SBM/TC, both yielding the same
    $B=11$ communities. The SBM yields a description length of
    $\Sigma_{\text{SBM}}=1761.1$ nats and the SBM/TC,
    $\Sigma_{\text{SBM/TC}}=1767.6$ nats.\label{fig:football}}
\end{figure}

\begin{figure}
  \includegraphics[width=\columnwidth]{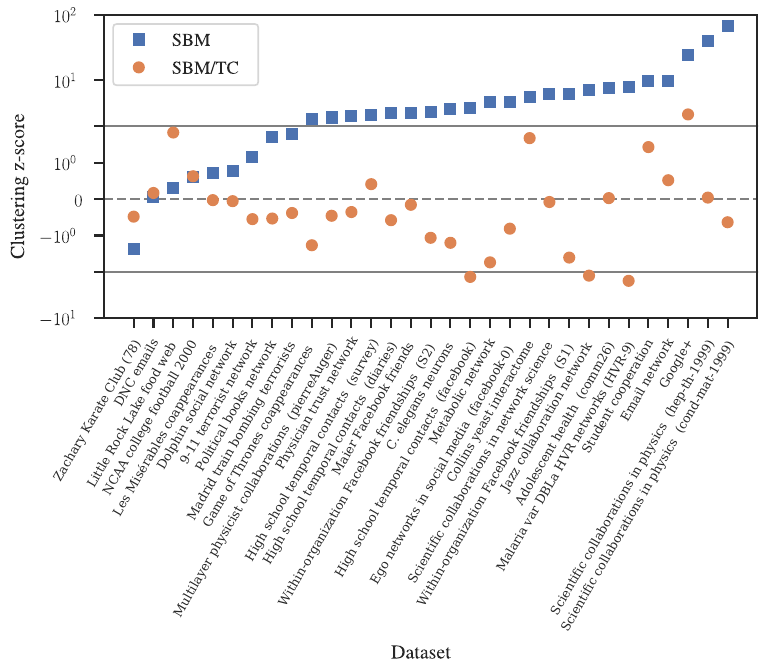}
  \includegraphics[width=\columnwidth]{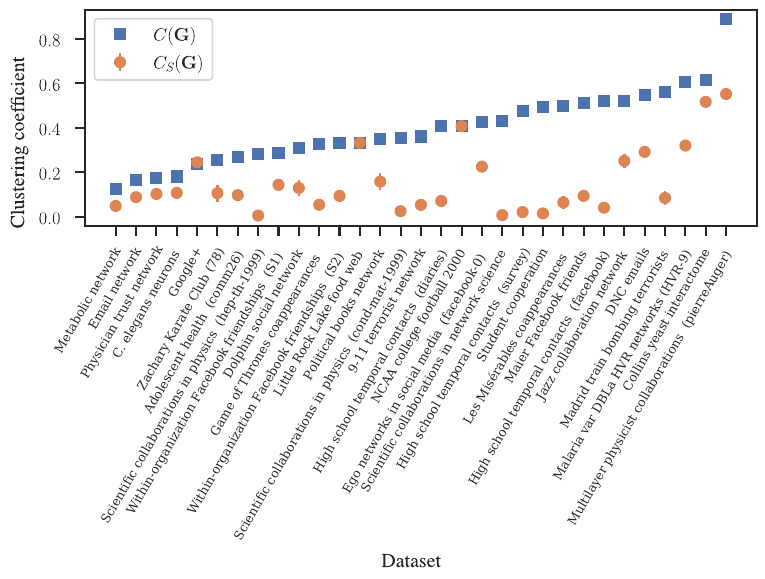}
  \includegraphics[width=\columnwidth]{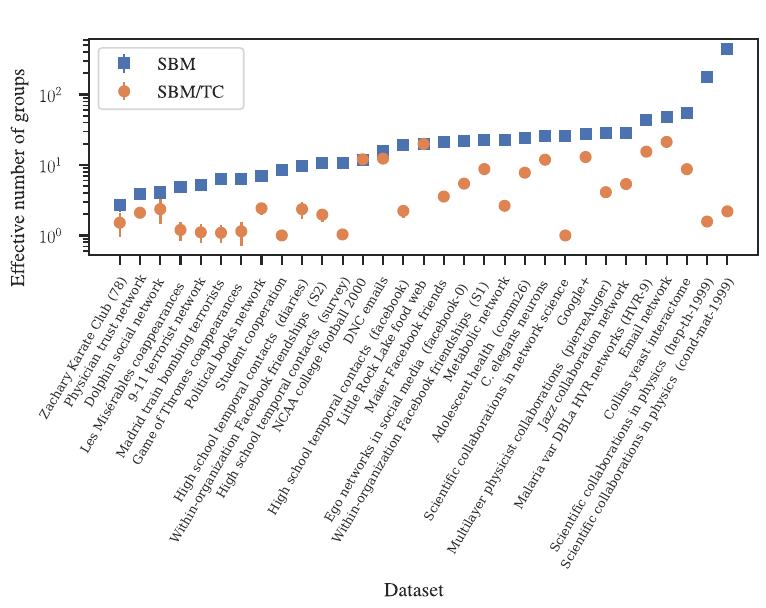}

  \caption{(top) Values of the z-score for the posterior predictive
  distributions of the clustering coefficient, as described in the text,
  for the SBM and SBM/TC as indicated in the legend, for different
  datasets. The solid horizontal lines mark the values $-2$ and
  $2$. (middle) Values of the clustering coefficient
  (Eq.~\ref{eq:clustering}) computed for the original network, $C(\G)$,
  and for the inferred seminal network, $C_S(\G)$, averaged over the
  posterior distribution according to Eq.~\ref{eq:c_A}, as shown in the
  legend. (bottom) Values of effective number of
  inferred groups, as given by Eq.~\ref{eq:eB}, for the SBM and SBM/TC
  as indicated in the legend. \label{fig:zscore}}
\end{figure}

We investigate the use our method with a variety of empirical
networks. We begin with a network of cooperation among students while
doing their homework for a course at Ben-Gurion
University~\cite{Fire_2012}. In Fig.~\ref{fig:student_cooperation}a we
show the network and a fit of the DC-SBM, which finds 9 assortative
communities. Based on this result --- and knowing that the partitions
found by inferring the SBM as we do here point to statistically
significant results that cannot be attributed to mere random
fluctuations~\cite{peixoto_bayesian_2019} --- we would be tempted to
posit that these divisions are uncovering latent social parameters that
could explain the observed cooperation between these groups of
students. However, if we employ instead the SBM/TC, we obtain the
result shown in Fig.~\ref{fig:student_cooperation}b, which uncovers
instead only a single group, and an abundance of triadic closure
edges. This is not unlike the artificial example considered in
Fig.~\ref{fig:random}, and points to a very different interpretation,
namely there is no measurable \emph{a priori} predisposition for
students to work with each other in groups, and the resulting network
stems instead from students choosing to work together if they already
share a mutual partner. Indeed if we inspect the description lengths
obtained with each model, we immediately recognize the SBM/TC as the
most plausible explanation, and therefore we deem the community
structure found by the SBM as an unlikely one by comparison.

We move now to another social network, but this time of friendships
between high school students~\cite{Moody_2001}. We show the results of
our analysis in Fig.~\ref{fig:netsci_add_health}. Using the SBM we find
$B=26$ groups, shown in Fig.~\ref{fig:netsci_add_health}a, which at
first seems like a reasonable explanation for this network. But instead,
with the SBM/TC we find only $B=9$ groups and a substantial amount of
triadic closure edges, as seen in
Fig.~\ref{fig:netsci_add_health}b. Differently from the previous
example, the SBM/TC still finds enough evidence for a substantial amount
of community structure, although with fewer groups than the pure
SBM. The groups found with the SBM/TC have a strong correlation with the
student grades, as shown in Fig.~\ref{fig:netsci_add_health}b, except
for the 11th and 12th grades, which seem to intermingle more, and for
which the model finds evidence of more detailed internal social
structures. This indicates that most of the subdivisions of the grades
found by the pure SBM are in fact better explained by triadic closure
edges, and the \emph{a priori} friendship preference within these grades
are far more homogeneous than the SBM fit would lead us to conclude. One
particularly striking feature of this analysis is that it imputes some
seemingly clear communities entirely to triadic closure. A good example
is the group highlighted with an arrow in
Fig.~\ref{fig:netsci_add_health}a, formed by students in the 8th
grade. According to the SBM/TC, this group has arisen due to the
formation of triangles between an initially poorly connected subset of
students, formed by all friends of a single student, rather than an
initial affinity between them. Comparing the SBM and the SBM/TC models,
we see that the latter has a substantially smaller description length
value, and hence needs considerably less information to place all the
edges in the network. We emphasize that this criterion takes into
account not only the likelihood of the respective model but also on its
complexity. In view of this, the SBM/TC hypothesis is objectively more
parsimonious, and in the absence of further data should be considered
more plausible than the pure SBM.

We move now to an additional example, this time of collaborations
between researchers in network science~\cite{Newman_2006}, shown in
Fig.~\ref{fig:netsci_add_health}. For this network, the SBM finds $B=27$
communities. The interpretation here is the same as previous analyses of
the same network, namely that these communities are groups that tend to
work together, with the occasional collaboration across groups. On the
other hand, when we employ the SBM/TC, the difference this time is quite
striking. Most of the community structure found with the pure SBM
vanishes and is replaced by a substrate network with a substantial
``core-periphery'' mixing pattern formed of two main groups, where the
``core'' (blue nodes) is composed of perceived initiators of the
collaborations with the ``periphery'' (yellow nodes), which end up being
connected in the final network simply by virtue of the all-to-all nature
of multi-way collaborations, captured here by triadic closure edges. The
core-periphery pattern is not perfect, as we observe seminal edges
between nodes of every type, but most commonly these exist between core
and periphery nodes, and the core nodes themselves, who therefore seem
to have a predisposition to wider collaborations. The difference between
the description lengths of both models is substantial, indicating that
the SBM/TC interpretation is indeed far more plausible.

Lastly, we consider the network of American football games between
colleges during the fall of 2000~\cite{Girvan_2002}, shown in
Fig.~\ref{fig:football}. For this network we observe an interesting
result, namely the SBM and SBM/TC yield the exact same inference,
corresponding very closely to the known division of the teams into
``conferences'' that tend to play with each other more frequently, which
means that SBM/TC gives a negligible probability of triadic closure
edges. Although we might expect this to occur for a network that has
very few or no triangles, and therefore substantial evidence against
triadic closure, this is not the case for the particular network in
question, which has in fact an abundance of triangles, in addition to
clear assortative communities. The reason for this is that, in this
particular case, the SBM is fully capable of accounting for the
triangles observed, which therefore can be characterized being a
``side-effect'' of the homophily between nodes of the same group,
instead of an excess that needs additional explanation. We will revisit
this particular case in the following, from a different angle.

One natural criticism of the SBM as a useful hypothesis for real
networks, however stylized as it clearly is, is that it assumes that
edges are placed independently with probability $O(B/N)$, for a network
with $N$ nodes and $B$ groups, assuming the group affinities are uniform
for all groups. One consequence of this is that the probability of
observing a spontaneous triadic closure edge will also scale with
$O(B/N)$. Therefore if $B \ll N$, we should not expect any abundance of
triangles, which is at odds with what we observe in many empirical
data. One problem with this logic is that we do not know \emph{a priori}
the precise relationship between $B$ and $N$ for finite empirical
networks, and therefore we cannot rule out the SBM hypothesis based
simply on an observed abundance of triangles. Auspiciously, with the
SBM/TC at hand, we are the perfect position to evaluate the SBM in that
regard, and understand how many of the observed triangles can be
attributed to an incidental link placement due to community structure,
or if they are instead better explained by explicit triadic closure
edges. A common way of quantifying the amount of triangles in a network
$\G$ is via its clustering coefficient $C(\G)\in[0,1]$, which determines
the fraction of triads in the network which are closed in a triangle,
and is given by
\begin{equation}\label{eq:clustering}
  C(\G) = \frac{\sum_{ijk}G_{ij}G_{jk}G_{ki}}{\sum_ik_i(k_i-1)},
\end{equation}
where $k_i=\sum_jG_{ij}$ is the degree of node $i$.
A meaningful way to evaluate whether a given model $P(\G|\bm\theta)$
with parameters $\bm\theta$ can capture what is seen in the data is to
compute the posterior predictive distribution,
\begin{equation}
  P(C|\G) = \sum_{\G'}\delta(C-C(\G'))\sum_{\bm\theta}P(\G'|\bm\theta)P(\bm\theta|\G).
\end{equation}
This involves sampling parameters $\bm\theta$ from the posterior
$P(\bm\theta|\G)$, generating new networks $\G'$ from the model
$P(\G'|\bm\theta)$, and obtaining the resulting population of $C(\G')$
values, which can then be compared to the observed value $C(\G)$, and in
this way we can determine if the model used is capable of capturing this
aspect of the data. In Fig.~\ref{fig:c-pred} we show the results of this
comparison for the SBM and SBM/TC (in Appendix~\ref{app:predictive} we
give more details about how $\bm\theta$ should be chosen in each case)
using three datasets. For three of the four networks we observe what one
might expect: although the SBM is capable of accounting for a
substantial amount of triangles (far more than one would expect by
naively assuming $B\ll N$), it falls short of explaining what is
actually seen in the data. The SBM/TC, on the other hand, accounts for a
realm of possibilities that comfortably includes what is observed in the
data, with a sufficiently high probability. For the remaining network in
Fig.~\ref{fig:c-pred}c, NCAA college football 2000, as before, we
observe a different picture. Namely, both models produce predictive
posterior distributions that are essentially identical, and fully
compatible with what is seen in the data. Therefore we can say with a
fair amount of confidence that the fairly high clustering coefficient
observed for this network can in principle be attributed to community
structure alone, rather than triadic closure, contradicting the
intuition obtained from the asymptotic case where $B\ll N$, which is not
applicable to this network.

We take the opportunity to emphasize that the results of
Fig.~\ref{fig:c-pred} demonstrate how the SBM/TC model is significantly
more well-behaved than ERGMs designed to reproduce triangle countes via
a maximum-entropy formulation~\cite{strauss_model_1975}. As demonstrated
in~\cite{park_statistical_2004, park_solution_2004,
foster_clustering_2011, fischer_sampling_2015}, these models define
ensembles with strong degeneracies, with most sampled networks having
either very low or very high triangle counts, but none with values
similar to what is actually seen in the modelled networks. This is not a
phenomenon we observe with the SBM/TC, where the clustering coefficient
distributions are unimodal, and concentrated on the empirical values.

We extend the previous analysis to a larger set of empirical networks,
as shown in Fig.~\ref{fig:zscore}, by summarizing the compatibility of
the posterior predictive distribution via the z-score,
\begin{equation}
  z = \frac{C(\G)-\avg{C}}{\sigma_C},
\end{equation}
where $\avg{C}$ and $\sigma_C$ are the mean and standard deviation of
the posterior predictive distribution. As we can see, there are a number
of networks for which the z-score values lie in the plausible interval
$[-2,2]$ for both models, but there is a much larger fraction of the
data for which the values for the SBM point to a decisive
incompatibility, whereas the SBM/TC yields credible values more
systematically.

We can further exploit the decomposition that the SBM/TC provides by
quantifying precisely, for any given network, how much of the observed
clustering can be attributed to triadic closure directly, or to
community structure indirectly. We can do so by computing the mean
clustering coefficient of the substrate seminal network from the
posterior distribution,
\begin{equation}\label{eq:c_A}
  C_S(\G) = \sum_{\{\A^{(l)}\},\{\g^{(l)}\},\bb}C(\A)P(\{\A^{(l)}\},\{\g^{(l)}\},\bb|\G).
\end{equation}
We can then compare this value with the coefficient for the observed
network $C(\G)$, as we show in Fig.~\ref{fig:zscore}. We identify a variety
of scenarios, including situations where the seminal network (and hence
the community structure) accounts for the majority of the observed
clustering, but most commonly we observe that a substantial fraction can
be attributed to more direct triadic closure. Nevertheless, in many
cases the values of $C_S(\G)$ do not drop to negligible values, showing
that the presence of triangles cannot be wholly attributed to either
mechanism in these cases. Indeed, this variability seems to indicate
that mere presence of a high or low density of triangles, as captured by
the clustering coefficient, cannot be used by itself to evaluate whether
triadic closure or community structure is the leading underlying
mechanism of network formation.

Another aspect of the suitability of triadic closure as a more plausible
network model is that it tends to come together with a less pronounced
inferred community structure, since part of the density heterogeneity
found is attributed to the former mechanism, rather than the latter. In
Fig.~\ref{fig:zscore} we characterize this difference by the effective
number of groups found with both models. We see that the discrepancy
between them is once again quite varied, where in some cases it can be
quite small, indicating that triadic closure plays a minor role, while
in other cases it can be quite extreme, indicating the dominant role
that triadic role has in the respective networks.

Overall, what we seem to extract from these empirical networks is that,
in the majority of cases (though not all), the observed structure seems
to be better explained by a heterogeneous combination of underlying
mixing patterns with a further distortion by an additional tendency of
forming triangles. The precise balance between these two components vary
considerably in general, and needs to be assessed for each individual
network.

\section{Edge prediction and network reconstruction}\label{sec:prediction}

\begin{figure}
  \begin{tabular}{c}
    Student cooperation\\
    \includegraphics[width=\columnwidth]{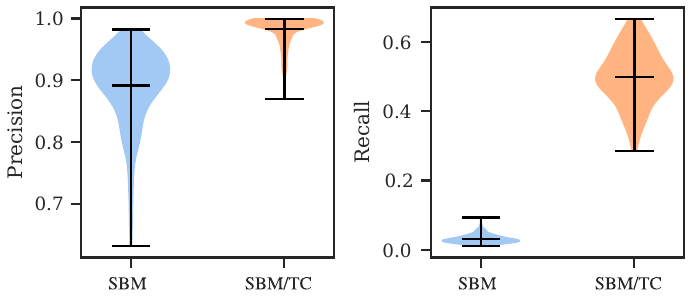}\\
    Scientific collaborations in Network Science\\
    \includegraphics[width=\columnwidth]{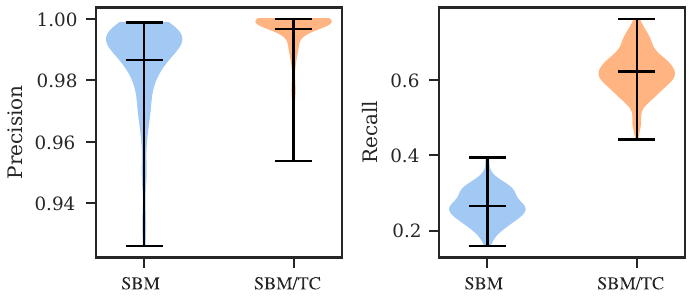}\\
    Adolescent health (comm26)\\
    \includegraphics[width=\columnwidth]{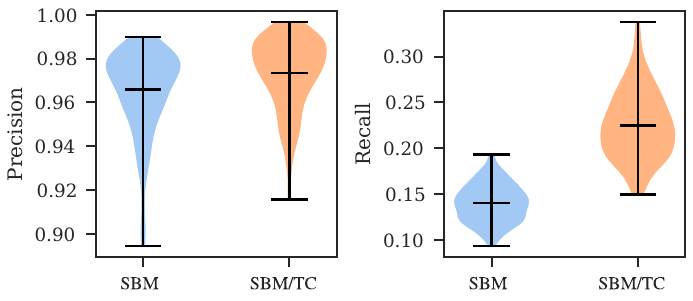}\\
    NCAA college football 2000\\
    \includegraphics[width=\columnwidth]{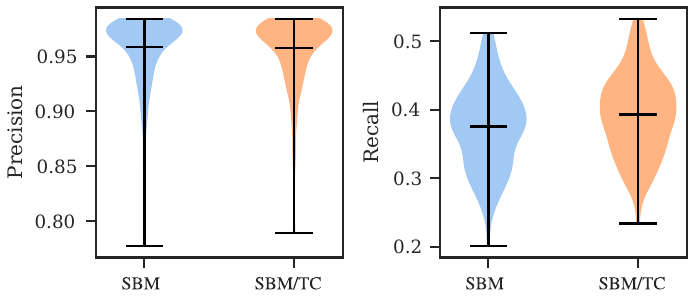}
  \end{tabular} \caption{Distributions of Precision and Recall values,
    according to the SBM and SBM/TC model, for four empirical networks,
    and a
    fraction $f=0.05$ of omitted edges and corresponding number of omitted
    non-edges. The results were obtained for $200$ different realizations
    of missing edges and non-edges.\label{fig:prediction}}
\end{figure}

As with every kind of empirical assessment, network data is subject to
measurement errors or omissions. A common use of network models is to
predict such erroneous and missing information from what is more
precisely known~\cite{clauset_hierarchical_2008,
guimera_missing_2009}. The SBM has been successfully used as such a
model~\cite{guimera_missing_2009, peixoto_reconstructing_2018}, since
the latent group assignments and the affinities between them can be
learned from partial network information, which in turn can be used to
infer what has been distorted or left unobserved. Another common
approach to edge prediction consists of attributing a higher probability
to a potential edge if it happens to form a
triangle~\cite{adamic_friends_2003}. As we have been discussing in this
work, these two properties --- group affinity and triadic closure ---
point to related but distinct processes of edge formation, and
approaches of edge prediction that rely exclusively on either one will
be maximally efficient only if it happens to be the dominant underlying
mechanism, which, as we have seen in the last section, is typically not
the case. However, with the SBM/TC model we have introduced, it should
be possible to accommodate both mechanisms at the same time, and in this
way improve edge prediction is more realistic settings. In the
following, we show how this can be done, and demonstrate it with a few
examples.

The scenario we consider is the general one presented in
Ref.~\cite{peixoto_reconstructing_2018}, where we make $n_{ij}$
measurements of node pair $(i,j)$ and record the number of times
$x_{ij}$ that an edge has been observed. Based on this data, we infer
the underlying network $\G$ according to the posterior distribution
\begin{equation}
  P(\G | \bm{n}, \bm{x}) = \frac{P(\bm{x}|\G,\bm{n})P(\G)}{P(\bm{x}|\bm{n})},
\end{equation}
with $\bm{n}=\{n_{ij}\}$ and $\bm{x}=\{x_{ij}\}$.  The measurement model
corresponds to a situation where the probabilities of observing missing
and spurious edges, $p$ and $q$ respectively, are uniform, leading to
\begin{multline}
  P(\bm{x}|\G,\bm{n},p,q) =
  \prod_{i<j}{n_{ij}\choose x_{ij}}\left[(1-p)^{x_{ij}}p^{n_{ij}-x_{ij}}\right]^{G_{ij}}\times\\
  \left[q^{x_{ij}}(1-q)^{n_{ij}-x_{ij}}\right]^{1-G_{ij}}.
\end{multline}
Assuming that both $p$ and $q$ are unknown a priori, i.e. $P(p)=P(q)=1$,
amounts to the marginal probability~\cite{peixoto_reconstructing_2018}
\begin{align}
  P(\bm{x}|\G,\bm{n})
  &= \int P(\bm{x}|\G,\bm{n},p,q)P(p)P(q)\;\dd p\,\dd q \\
  & = \left[\prod_{i<j}{n_{ij}\choose x_{ij}}\right] {\E \choose \T}^{-1}\frac{1}{\E+1}\times\nonumber\\
  &\qquad
  {\M-\E \choose \X-\T}^{-1}\frac{1}{\M-\E+1}.
\end{align}
where we have
\begin{align}
  \M&=\sum_{i<j}n_{ij},  &  \X&=\sum_{i<j}x_{ij},\\
  \E&=\sum_{i<j}n_{ij}G_{ij}, &  \T&=\sum_{i<j}x_{ij}G_{ij}.
\end{align}
The network model comes into play via the prior $P(\G)$. For the SBM/TC model this is
\begin{equation}
  P(\G) = \sum_{\{\g^{(l)}\}, \{\A^{(l)}\}, \bb}P(\G, \{\g^{(l)}\}, \{\A^{(l)}\}, \bb).
\end{equation}
Once more, we avoid an intractable computation, by sampling instead from
a joint posterior with the model parameters, i.e.
\begin{multline}
  P(\G, \{\g^{(l)}\}, \{\A^{(l)}\}, \bb | \bm{n}, \bm{x}) =\\
  \frac{P(\bm{x}|\G,\bm{n})P(\G, \{\g^{(l)}\}, \{\A^{(l)}\}, \bb)}{P(\bm{x}|\bm{n})}
\end{multline}
so that the desired posterior distribution can be obtained by
marginalization
\begin{equation}
  P(\G | \bm{n}, \bm{x}) = \sum_{\{\g^{(l)}\}, \{\A^{(l)}\}, \bb}P(\G, \{\g^{(l)}\}, \{\A^{(l)}\}, \bb | \bm{n}, \bm{x}).
\end{equation}
In order to perform our comparison, we consider the following particular
setup for the data $(\bm n,\bm x)$. Given a true network $\G$ we select
a random subset $\bm{P}_t$ of the edges (``true positives''), and an
equal-sized random subset $\bm{N}_t$ of ``non-edges'' (``true
negatives''), i.e., node pairs $(i,j)$ for which $G_{ij}=0$, such that
$|\bm{P}_t|=|\bm{N}_t|=f E$, where $f\in[0,1]$ is a free parameter and
$E$ is the total number of edges. We then set $n_{ij}\to\infty$ for all
node pairs neither in $\bm{P}_t$ nor in $\bm{N}_t$, with $x_{ij}=n_{ij}$
if $G_{ij}=1$ and $x_{ij}=0$ otherwise
--- these are parts of the network about which we are perfectly
certain. For the node pairs in $\bm{P}_t$ and $\bm{N}_t$ we set
$n_{ij}=x_{ij}=0$, meaning we lack any data about them. We then compute
the posterior marginal probability
\begin{equation}
  p_{ij} = \sum_{\G}G_{ij}P(\G|\bm n,\bm x),
\end{equation}
and we use it to evaluate the quality of the reconstruction. We do so by
computing the Precision and Recall, defined as
\begin{align}
  \text{Precision} &= \frac{\sum_{(i,j)\in
  \bm{P}_t}p_{ij}}{\sum_{(i,j)\in \bm{P}_t\cup\bm{N}_t}p_{ij}}\\
  \text{Recall} &= \frac{\sum_{(i,j)\in
      \bm{P}_t}p_{ij}}{|\bm{P}_t|},
\end{align}
which measures the fraction of correctly predicted edges, relative to
the total number of edges predicted, or the total number of true edges,
respectively.

In Fig.~\ref{fig:prediction} we show the results of the above analysis
for some of the networks studied previously, using both the SBM/TC model
and the pure SBM. For most of them, the SBM/TC model yields a superior
predictive performance, sometimes substantially. This shows that while
community detection via the SBM can to some extent detect the patterns
induced by triadic closure, the more explicit SBM/TC model does a better
job at this, corroborating the model selection arguments we have used
previously. For networks of games between American football college
teams, the situation is once again different, and we observe
indistinguishable results between the SBM and SBM/TC. For this network,
as the previous analysis has established, triadic closure seems to play
a insignificant role, despite the relative abundance of triangles. As a
consequence, in this case the SBM/TC model offers no advantage in edge
prediction, but importantly, it does not degrade it either.

In a recent work, Ghasemian \etal~\cite{ghasemian_stacking_2020} have
performed a large-scale analysis of over two hundred edge prediction
methods on over five hundred networks belonging to various
domains. Although the overall conclusion of that work was that no single
method dominates on every data, the predictive performance of the
different methods were far from uniform, with the method above based on
the SBM providing the single best performance
overall.\footnote{Ghasemian \etal~\cite{ghasemian_stacking_2020}
considered only a simplified version of the method described, where only
the best-scoring partition was used, instead of an average over the
posterior distribution. Furthermore, they have used only the version of
the SBM with noninformative priors, which is known to underfit, as
opposed to the nested SBM~\cite{peixoto_hierarchical_2014,
peixoto_nonparametric_2017} which removes this problem. Accounting for
both of these issues have been shown before to improve edge prediction
systematically~\cite{valles-catala_consistencies_2018}, and could
potentially have pushed the result of the analysis in
Ref.~\cite{ghasemian_stacking_2020} even more in favor of the SBM
approach.} Interestingly, the situations where the SBM approach yielded
inferior performance were precisely for social networks, for which some
predictors based on triadic closure performed better. Although our
results above fall short of a thorough and systematic analysis of the
wide domains of network data, since we consider only a handful of
networks, they nevertheless seem to give good indication that combining
group affinity with triadic closure could potentially eliminate this
shortcoming for this particular class of network data.

\section{Discussion}\label{sec:conclusion}

We have presented a generative model and corresponding inference scheme
that is capable of differentiating community structure from triadic
closure in empirical networks. We have shown that although these
features are typically conflated in traditional network analysis, our
method can pick them apart, allowing us tell us whether an observed
abundance of triangles is a byproduct of an underlying homophily between
nodes, or whether they arise out of a local property of
transitivity. Likewise, we have also shown how our method can evade the
detection of spurious communities, which are not due to homophily, but
arise instead simply out of a random formation of triangles.

Our approach shows how local and global (or mesoscale) generative
processes can be combined into a single model. Since it contains a
mixture of both mechanisms, our method is able to decompose them for a
given observed network according to their inferred contributions. By
employing our method on several empirical networks, we were able to
demonstrate a wide variety of scenarios, containing everything from a
large number of triangles caused predominantly by triadic closures, by a
mixture of community structure and triadic closures, and by community
structure alone. These findings seem to indicate that local and global
network properties tend to mix in nontrivial ways, and we should refrain
from automatically concluding that an observed local property
(e.g. large number of triangles) cannot have a global cause (e.g. group
homophily), and likewise an observed global property (e.g. community
structure) cannot have a purely local cause (e.g. triadic closure). Our
explicit mixture approach could in principle be extended also to other
types of local structures such as reciprocity in directed
networks~\cite{safdari_generative_2020}, or higher-order motifs,
bringing further insights into how these local properties are entangled
with global ones.

Several authors had shown before that triadic closure can induce the
formation of community structure in
networks~\cite{foster_clustering_2011,foster_communities_2010,lopez_transitions_2020,bianconi_triadic_2014,wharrie_micro-_2019,
asikainen_cumulative_2020}. This introduces a problem of interpretation
for community detection methods that do not account for this, which, to
the best of our knowledge, happens to be the vast majority of them. This
is true also for inference methods based on the SBM, which, although
they are not susceptible to finding spurious communities formed by a
fully random placement of edges~\cite{guimera_modularity_2004} (unlike
non-inferential methods, which tend to
overfit~\cite{ghasemian_evaluating_2019}),\footnote{Overfitting here
means that the number of communities found is too large, and that the
method can even find communities in completely random networks.} they
cannot evade those arising from triadic
closure~\cite{wharrie_micro-_2019}. Our approach provides a solution to
this interpretation problem, allowing us to reliably rule out triadic
closure when identifying communities in networks.

We have also shown how incorporating triadic closure together with
community structure can improve edge prediction, without degrading the
performance in situations where it is not present. This further
demonstrates the usefulness of approaches that model networks in
multiple scales, combining multiple edge generation mechanisms, and
points to a general way of systematically improving our understanding of
network data.

\bibliography{bib,data}

\appendix

\section{Latent multigraph SBM} \label{app:multigraph}

The marginal likelihood of Eq.~\ref{eq:sbm} is in fact obtained for a
multigraph model~\cite{peixoto_nonparametric_2017}, where the adjacency
entries can take any natural value, $A_{ij}\in\mathbf{N}$. Although we
could in principle ignore this discrepancy, since this kind of model
generates simple graphs as a special case, this comes at the expense of
a reduced expressiveness of the model~\cite{peixoto_latent_2020}, since
this kind of multigraph model cannot describe the placement of single
edges with high probability, or account for the emergent degree-degree
correlations that must be present in simple graphs. Instead, here we
take the approach proposed
in~\cite{peixoto_reconstructing_2018,peixoto_latent_2020}, and consider
a \emph{latent} multigraph $\A'$, with $\A'_{ij}\in\mathbf{N}$, which is
then converted into a simple graph $\A(\A')$ simply by ignoring the edge
multiplicities, i.e.
\begin{equation}
  A_{ij} =
  \begin{cases}
    1 & \text{ if } A_{ij}' > 0,\\
    0 & \text{ otherwise.}
  \end{cases}
\end{equation}
The latent multigraph $\A'$ is generated according to Eq.~\ref{eq:sbm},
which means the simple graph $\A$ is generated according to
\begin{equation}
  P(\A|\bb) = \sum_{\A'}\bm{1}_{\{\A(\A')=\A\}}P(\A'|\bb).
\end{equation}
Instead of working with this marginal probability directly (which is
intractable), we infer the latent edge multiplicities as well, from a
joint posterior distribution
\begin{equation}
  P(\g,\A',\bb | \G) = \frac{P(\G,\g|\A(\A'))P(\A'|\bb)P(\bb)}{P(\G)},
\end{equation}
where the simple graph $\A(\A')$ is used for the triadic closure
likelihood $P(\G,\g|\A)$. In this way, the inference procedure is the
same as the one described in the main text, with the only modification
that we need to infer the integer values of $\A'$ rather than its binary
values.

\section{Expected density of transitivity} \label{app:general}

As mentioned in the main text, the choice of priors used for
Eq.~\ref{eq:marginal_simple} makes the calculation very simple, but it
implies that we expect the observed graphs to always have a large
fraction of triadic closures. An outcome of this is that the probability
of observing a final graph without any triadic closure,
i.e. $\sum_{uij}g'_{ij}(u)=0$, is given by
\begin{align}
  P(\g'|\A)&=\prod_u\left[1+\sum_{i<j}m_{ij}(u)\right]^{-1}\\
  &=O\left(\frac{1}{[\avg{k^2}-\avg{k}]^N}\right),
\end{align}
which is exponentially suppressed for a large number of nodes $N$. Even
though we are interested in modeling networks which do posses some
amount of triadic closure, we should be \emph{a priori} more agnostic
about the actual amount, as to also accommodate situations where this
property is not abundant. We can address this by noting that the
likelihood of Eq.~\ref{eq:marginal_simple} can be alternatively
interpreted as the one of a fully equivalent model given by
\begin{equation}
  P(\g|\A) = \prod_u\sum_{E_u} P(\g(u)|\A, E_u)P(E_u|\A),
\end{equation}
where
\begin{equation}
  P(\g(u)|\A, E_u) = \frac{\bm{1}_{\{E_u = \sum_{i<j}g_{ij}(u)\}}}{
  {\sum_{i<j}m_{ij}(u) \choose E_u}}
\end{equation}
is the probability of uniformly sampling an ego graph $\g(u)$ with
exactly $E_u=\sum_{i<j}g_{ij}(u)$ edges, and
\begin{equation}
P(E_u|\A) = \frac{1}{1+\sum_{i<j}m_{ij}(u)}
\end{equation}
is the probability of uniformly sampling the number of edges in $\g(u)$
in the allowed range $[0,\sum_{i<j}m_{ij}(u)]$. This
interpretation allows us to do a small modification of our model that
makes it more versatile, namely we separate the nodes into two sets,
according to an auxiliary binary variable $t_u \in \{0,1\}$, such that
if $t_u=0$ then the corresponding ego graph has no edges,
$P(E_u|\A,t_u=0)=\delta_{0,E_u}$, otherwise it has a nonzero number of
edges, sampled uniformly as
\begin{equation}
  P(E_u|\A,t_u=1) = \frac{1-\delta_{E_u,0}}{\sum_{i<j}m_{ij}(u)}.
\end{equation}
The modified marginal distribution becomes then
\begin{widetext}
\begin{align}
  P(\g|\A) &= \sum_{\bm t}\left[\prod_u\sum_{E_u} P(\g(u)|\A, E_u)P(E_u|\A,\bm t)\right]P(\bm t|\A)\\
  &=\left\{\prod_u\left[{\sum_{i<j}m_{ij}(u) \choose \sum_{i<j}g_{ij}(u)}\sum_{i<j}m_{ij}(u)\right]^{\delta_{1,\Theta\left[\sum_{i<j}g_{ij}(u)\right]}}\right\}
  {\sum_u\Theta\left[\sum_{i<j}m_{ij}(u)\right]\choose \sum_u\Theta\left[\sum_{i<j}g_{ij}(u)\right]}^{-1}\times
  \frac{1}{1+\sum_u\Theta\left[\sum_{i<j}m_{ij}(u)\right]},\label{eq:marginal}
\end{align}
\end{widetext}
where $\Theta[x] = \{1 \text{ if } x > 0, \text{ else } 0\}$ is the
Heaviside step function, and we have used the prior
\begin{equation}
  P(\bm t| \A) = \sum_{N_t}P(\bm t|\A,N_t)P(N_t|\A),
\end{equation}
with
\begin{equation}
  P(\bm t|\A,N_t) = \frac{\bm{1}_{\{\sum_u t_u = N_t\}}\prod_u\Theta\left[\sum_{i<j}m_{ij}(u)\right]^{t_u}}{{\sum_u\Theta\left[\sum_{i<j}m_{ij}(u)\right] \choose N_t}},
\end{equation}
and
\begin{equation}
  P(N_t|\A) = \frac{1}{1+\sum_u\Theta\left[\sum_{i<j}m_{ij}(u)\right]}.
\end{equation}
The above amounts to sampling in sequence the number of nodes $N_{t}$
and the partition $\bm{t}$, both uniformly at random from the allowed
range. Although these equations take longer to write, they are not much
more difficult to use. As a result of this parametrization, if we
consider again the particular graph with no triadic closures,
i.e. $\sum_{uij}g_{ij}'(u)=0$, it is generated with probability
\begin{equation}
  P(\g'|\A)=\frac{1}{1+\sum_u\Theta[\sum_{i<j}m_{ij}(u)]}
  =O\left(\frac{1}{N}\right),
\end{equation}
which is relatively large and no longer exponentially suppressed for
large $N$, meaning that our model can also accommodate the same kinds of
networks that are sampled from the pure SBM, without triadic
closures. This does not mean that the modified model generates
\emph{typical} networks with substantially smaller number of transitive
edges, only that the variance with respect to this property is larger,
and the model is thus more indifferent about the potential networks that
are possible to be observed.

As mentioned in the main text, this modification makes the SBM fully
nested inside the SBM/TC, as we have
\begin{align}
  P(\G,\g',\A=\G|\bb) &= \frac{P(\G|\bb)}{1+\sum_u\Theta\left[\sum_{i<j}m_{ij}(u)\right]}\\
  &\ge \frac{P(\G|\bb)}{N+1},
\end{align}
with $\g'$ being empty ego graphs, and the last equality achieved if
$\sum_{i<j}m_{ij}(u) > 0$ for every node $u$.

\subsection{Iterated triadic closure}

For the generalized model with iterated triadic closures, the marginal
likelihood is also analogous to Eq.~\ref{eq:marginal},
\begin{widetext}
\begin{multline}
  P(\g^{(l)}|\A^{(l-1)}, \g^{(l-1)})
  =\\
  \left\{\prod_u\left[{\sum_{i<j}m_{ij}^{(l)}(u) \choose \sum_{i<j}g_{ij}^{(l)}(u)}\sum_{i<j}m_{ij}^{(l)}(u)\right]^{\delta_{0,\sum_{i<j}g_{ij}^{(l)}(u)}-1}\right\}
  {\sum_u\bm{1}_{\{\sum_{i<j}m_{ij}^{(l)}(u)>0\}}\choose \sum_u\bm{1}_{\{\sum_{i<j}g^{(l)}_{ij}(u) > 0\}}}^{-1}
  \frac{1}{1+\sum_u\bm{1}_{\{\sum_{i<j}m_{ij}^{(l)}(u)>0\}}}.
\end{multline}
\end{widetext}

\section{MCMC moves} \label{app:mcmc}

The MCMC algorithm described in the main text is implemented with the
following moves. The first kind is to attempt to move an edge $(i,j)$ in
ego graph $\g^{(l)}(u)$ at its current generation $l\in[0,L]$ to another
ego graph $\g^{(l')}(v)$ for $v\neq u$ at generation $l'\ne l$.  We do so
by selecting first an edge $(i,j)$ in $\G$ as well as a generation $l$,
both uniformly at random, and an ego node $u$ that is relevant to edge
$(i,j)$ at generation $l$, also uniformly at random. The number of ego
graphs that are relevant for this edge is given by
\begin{equation}
  n_{ij}^{(l)} = \sum_uA^{(l-1)}_{ui}(1-A^{(l-1)}_{uj}),
\end{equation}
which is independent on the value of $g_{ij}^{(l')}(u)$ for any $l'$. We
then sample another generation $l'\ne l$ and proceed in the same way to
sample a relevant ego node $v$. In either case, if $l=0$ is selected,
then the choice of an ego graph is not made, since we are selecting
simply an entry $(i,j)$ in $\A$ with probability one. The final
probability of selecting the move $(i,j,u,l) \to (i,j,v,l')$, assuming
$l>0$ and $l'>0$, is given by
\begin{equation}
  P(i,j,u,v,l,l'|\{\g^{(l)}\},\{\A^{(l)}\})=
  \frac{1}{E_Gn_{ij}^{(l)}n_{ij}^{(l')}L(L+1)},
\end{equation}
where $E_G$ is the number of edges in $\G$. Given this selection we then
make the change $g_{ij}'^{(l)}(u)=g_{ij}^{(l)}-1$ and
$g_{ij}'^{(l')}(u)=g_{ij}^{(l')}+1$, and accept it with probability
\begin{widetext}
  \begin{equation}
  \min\left(1,\frac{P(\{{\g'}^{(l)}\},\{{\A^{(l)}}'\},\bb | \G)P(i,j,u,v,l,l'|\{{\g}^{(l)}\},\{{\A^{(l)}}'\})}
           {P(\{\g^{(l)}\},\{\A^{(l)}\},\bb | \G)P(i,j,v,u,l',l|\{{\g'}^{(l)}\},\{\A^{(l)}\})}\right)
           = \min\left(1,\frac{P(\{{\g'}^{(l)}\},\{{\A^{(l)}}'\},\bb | \G)}
           {P(\{\g^{(l)}\},\{\A^{(l)}\},\bb | \G)}\right)
  \end{equation}
\end{widetext}
which is independent on the actual move probabilities, since they
remain the same after and before the move. Note that invalid moves that
result in $\g^{(l)}_{ij}<0$ or $\A^{(l)}_{ij}<0$ are always rejected in this way.

In addition, we also make a second kind of move by selecting again an
edge $(i,j)$ in $\G$ as well as a generation $l$, both uniformly at
random, and an ego node $u$ that is relevant to edge $(i,j)$ at
generation $l$, with the same probability as before. We then make the
move ${g'}_{ij}^{(l)} = g_{ij}^{(l)} \pm 1$ with probability $1/2$,
and accept again according to
\begin{equation}
  \min\left(1,\frac{P(\{{\g'}^{(l)}\},\{\A^{(l)}\},\bb | \G)}
           {P(\{\g^{(l)}\},\{\A^{(l)}\},\bb | \G)}\right).
\end{equation}
In case $l=0$ is selected, the move is different, due to the multigraph
nature of $\A$. We make instead the proposal $A_{ij}\to A_{ij}'$
according to a geometric distribution with mean $A_{ij}+1$,
\begin{equation}
  P(A_{ij}'|A_{ij}) = \left(\frac{A_{ij}+1}{A_{ij}+2}\right)^{A_{ij}'}\frac{1}{A_{ij}+2}.
\end{equation}
In this case, the acceptance probability changes to
\begin{equation}
  \min\left(1,\frac{P(\{{\g}^{(l)}\},\{{\A^{(l)}}'\},\bb | \G)P(A_{ij}|A_{ij}')}
           {P(\{\g^{(l)}\},\{\A^{(l)}\},\bb | \G)P(A_{ij}'|A_{ij})}\right).
\end{equation}

Finally, the last kind of move involves a change in partition
$\bb\to\bb'$ from the proposal $P(\bb'|\bb)$, which is accepted with
probability
\begin{equation}
  \min\left(1,
  \frac{P(\A|\bb')P(\bb')P(\bb|\bb')}
       {P(\A|\bb)P(\bb)P(\bb'|\bb)}\right).
\end{equation}
For the latter we use the merge-split moves, combined with single-node
moves, described in Ref.~\cite{peixoto_merge-split_2020}.

The moves above fulfill detailed balance, and when combined, they also
preserve ergodicity, since they allow every latent multigraph,
decomposition into ego graphs, and node partition to be sampled. Due to
this, with sufficiently many iterations the algorithm must eventually
produce samples from the desired posterior distribution.

\subsection{Algorithmic complexity}

We can break down the time complexity of the above algorithm as
follows. At any given time, we keep track of all relevant ego graphs for
each edge $(i,j)$ in $\G$, those that have edge $(i,j)$ in them, as well
as the number of edges $E_u^{(l)} = \sum_{i<j}g_{ij}^{(l)}(u)$ of every
ego graph. Based on this bookkeeping, whenever an entry
$g_{ij}^{(l)}(u)$ (or $A_{ij}$ if $l=0$) is modified, to compute the
log-likelihood difference we need only to evaluate the common neighbors
of $i$ and $j$ or the new or removed open and closed triads $(i,j,v)$ or
$(v,i,j)$ that affect generation $l+1$, both of which can be computed in
$O(k_i + k_j)$. As a result, a whole ``sweep'' of the MCMC algorithm,
where each edge in $\G$ had a chance to be moved by one of the proposals
considered, has an overall complexity of $O(N\avg{k^2})$, since each
node $i$ has $k_i$ edges that need to be moved at every sweep, each of
which requiring time $O(k_i+k_j)$, with $j$ being the other endpoint.

For the partition part of the algorithm, the overall complexity of a
sweep, where every node had a chance to be moved to a different group,
is $O(E + N)$, independent of the number of groups being
occupied~\cite{peixoto_merge-split_2020}.

Combining the two kinds of moves gives us an overall complexity of
$O(N\avg{k^2})$ per sweep, which for sparse graphs with $\avg{k^2}=O(1)$
amounts to $O(N)$. This means that it is possible, at least in
principle, to apply this algorithm for very large networks.

On top of the time it takes to perform sweeps of the MCMC, there is also
the mixing time of the Markov chain, which determines how long one needs
to wait before usable samples from the posterior distribution are
made. It is difficult to estimate the mixing time, as it depends heavily
on the actual network structure being considered, but we found that the
algorithm gives usable results in reasonable time even for networks with
over a hundred thousand to a million edges, although we did not attempt
a detailed investigation of networks which are much larger than this.

We have evaluated the quality of the algorithm with the analysis
presented in Fig.~\ref{fig:recovery}, where networks from the SBM/TC
model were generated, and the inference was performed in them. By
comparing the obtained results wit the true values of the latent
parameters, we observed that the triadic closure component was
identified with excellent accuracy, and the SBM component was identified
with an accuracy indistinguishable from when considering only the pure
SBM case, all the way down to the detectability transition. This gives
us a very good amount of confidence that the method converges, at least
in controlled scenarios.

When applied to empirical networks, the diagnostics performed were to
run the algorithm many times and evaluate if similar results are
produced, which happened to be the case with the data analyzed.

A reference implementation of this algorithm is freely made available as
part of the \texttt{graph-tool} library~\cite{peixoto_graph-tool_2014}.

\section{Predictive posterior distribution} \label{app:predictive}

The predictive posterior distribution considered in the main text is
\begin{equation}
  P(C|\G) = \sum_{\G'}\delta(C-C(\G'))\sum_{\bm\theta}P(\G'|\bm\theta)P(\bm\theta|\G),
\end{equation}
where $\bm\theta$ are the parameters of model $P(\G|\bm\theta)$. Here we
specify more precisely how these parameters are chosen and sampled for
the SBM/TC model. The marginal likelihood for the SBM given by
Eq.~\ref{eq:sbm} can be written equivalently
as~\cite{peixoto_nonparametric_2017}
\begin{equation}
  P(\A|\bb) = P(\A|\bm{k},\bm{e},\bb)P(\bm{k}|\bm{e},\bb)P(\bm{e}|\bb),
\end{equation}
where the likelihood of the microcanonical DC-SBM is given by
\begin{equation}
  P(\A|\bm{k},\bm{e},\bb)=
  \frac{\prod_{r<s}e_{rs}!\prod_re_{rr}!!\prod_ik_i!}
       {\prod_{i<j}A_{ij}!\prod_iA_{ii}!!\prod_re_r!},
\end{equation}
and prior for the degrees is
\begin{equation}
  P(\bm{k}|\bm{e},\bb)=
  \prod_r\frac{\prod_k\eta_k^r!}{n_r!q(e_r,n_r)},
\end{equation}
and the prior for the edge counts between groups is
\begin{equation}
  P(\bm{e}|\bb)={\frac{B(B+1)}{2} + E - 1 \choose E}^{-1}.
\end{equation}
For the triadic closure edges we have the likelihood $P(\g(u)|\A,p_u)$
of Eq.~\ref{eq:p_g}, which, given a uniform prior $P(p_u)=1$, gives us a
a Beta posterior distribution
\begin{multline}
  P(p_u|\g(u),\A) =\\
  \frac{p_u^{\sum_{i<j}g_{ij}(u)m_{ij}(u)}(1-p_u)^{\sum_{i<j}(1-g_{ij}(u))m_{ij}(u)}}
  {\mathcal{B}\left(\textstyle\sum_{i<j}g_{ij}(u)m_{ij}(u),\textstyle\sum_{i<j}(1-g_{ij}(u))m_{ij}(u)\right)},
\end{multline}
where $\mathcal{B}(x,y)$ is the Beta function. Based on this
parametrization, our predictive posterior distribution is obtained by
setting $\theta=(\{\bm p^{(l)}\}, \bm{k}, \bm{e}, \bb)$, amounting to
\begin{widetext}
\begin{multline}
  P(C|\G) = \sum_{\substack{\{{\g}^{(l)}\}\\\{{\g'}^{(l)}\}\\ \A, \A'\\ \bm{k}, \bm{e}, \bb}}\int\,d\{\bm{p}^{(l)}\} \delta\left\{C-C\left[\G(\A,\{\g^{(l)}\})\right]\right\} \left[\prod_{l,u} P(\g^{(l)}(u)|p_u^{(l)},\A)\right]P(\A|\bm{k}, \bm{e}, \bb)\times\\ \left[\prod_{l,u} P(p_u^{(l)}|{{\g'}^{(l)}(u)},\A')\right]
  P(\{{\g'}^{(l)}\}, \A',\bm{k}, \bm{e}, \bb|\G).
\end{multline}
\end{widetext}
Operationally, this just means running our inference algorithm to obtain
our latent variables $\{{\g'}^{(l)}\}, \{{\A^{(l)}}'\},\bm{k}, \bm{e}$ and $\bb$,
and the triadic closure propensities $\bm{p}^{(l)}$ from its
posterior, using that to obtain a new seminal network $\A$ from the same
SBM, together with its new ego graphs $\{\g^{(l)}\}$, and then finally
computing the resulting clustering coefficient.

\section{Network datasets}

Below are descriptions of the network datasets used in this work. The
codenames in parenthesis correspond to the respective entries in the
Netzschleuder repository~\cite{peixoto_netzschleuder_2020} where the
networks can be downloaded. Some of the descriptions were obtained from
the Colorado Index of Complex Networks~\cite{clauset_colorado_2016}.

\input{data-edited.tex}

\end{document}

%% file: data-edited.tex
\subparagraph{Adolescent health (\href{https://networks.skewed.de/net/add_health}{\texttt{add\_health}})~\cite{Moody_2001}:} A directed network of friendships obtained through a social survey of high school students in 1994. The ADD HEALTH data are constructed from the in-school questionnaire; 90,118 students representing 84 communities took this survey in 1994-95. Some communities had only one school; others had two. Where there are two schools in a community students from one school were allowed to name friends in the other, the ``sister school''. For this analysis, a  symmetrized version of the original directed network has been used, considering only its largest connected component. The particular network named \texttt{comm26} has been used. This network has $N=551$ nodes and $E=2624$ edges.
\subparagraph{Scientific collaborations in physics (\href{https://networks.skewed.de/net/arxiv_collab}{\texttt{arxiv\_collab}})~\cite{Newman_2001}:} Collaboration graphs for scientists, extracted from the Los Alamos e-Print arXiv (physics), for 1995-1999 for three categories, and additionally for 1995-2003 and 1995-2005 for one category. For copyright reasons, the MEDLINE (biomedical research) and NCSTRL (computer science) collaboration graphs from this paper are not publicly available. For this analysis, only the largest connected component of the networks were considered. The particular networks named \texttt{cond-mat-1999}, \texttt{hep-th-1999} have been used, with number of nodes and edges, $(N,E)$, given by (13861, 44619), (5835, 13815), respectively.
\subparagraph{Metabolic network (\href{https://networks.skewed.de/net/celegans_metabolic}{\texttt{celegans\_metabolic}})~\cite{Duch_2005}:} List of edges comprising the metabolic network of the nematode \emph{C. elegans}. This network has $N=453$ nodes and $E=4596$ edges.
\subparagraph{C. elegans neurons (\href{https://networks.skewed.de/net/celegansneural}{\texttt{celegansneural}})~\cite{1986,Watts_1998}:} A network representing the neural connections of the Caenorhabditis elegans nematode. For this analysis, a symmetrized version of the original directed network has been used. This network has $N=297$ nodes and $E=2359$ edges.
\subparagraph{Collins yeast interactome (\href{https://networks.skewed.de/net/collins_yeast}{\texttt{collins\_yeast}})~\cite{Collins_2007}:} Network of protein-protein interactions in Saccharomyces cerevisiae (budding yeast), measured by co-complex associations identified by high-throughput affinity purification and mass spectrometry (AP/MS). For this analysis, only the largest connected component of the network was considered. This network has $N=1004$ nodes and $E=8319$ edges.
\subparagraph{DNC emails (\href{https://networks.skewed.de/net/dnc}{\texttt{dnc}})~\cite{Kunegis_2013}:} A network representing the exchange of emails among members of the Democratic National Committee, in the email data leak released by WikiLeaks in 2016. For this analysis, only the largest connected component of the network was considered. This network has $N=849$ nodes and $E=12038$ edges.
\subparagraph{Dolphin social network (\href{https://networks.skewed.de/net/dolphins}{\texttt{dolphins}})~\cite{Lusseau_2003}:} An undirected social network of frequent associations observed among 62 dolphins (Tursiops) in a community living off Doubtful Sound, New Zealand, from 1994-2001. This network has $N=62$ nodes and $E=159$ edges.
\subparagraph{Ego networks in social media (\href{https://networks.skewed.de/net/ego_social}{\texttt{ego\_social}})~\cite{1210.8182v3}:} Ego networks associated with a set of accounts of three social media platforms (Facebook, Google+, and Twitter). Datasets include node features (profile metadata), circles, and ego networks, and were crawled from public sources in 2012. For this analysis, only the largest connected component of the network was considered. The particular network named \texttt{facebook\_0} has been used. This network has $N=324$ nodes and $E=2514$ edges.
\subparagraph{Maier Facebook friends (\href{https://networks.skewed.de/net/facebook_friends}{\texttt{facebook\_friends}})~\cite{1706.02356v2}:} A small anonymized Facebook ego network, from April 2014. Nodes are Facebook profiles, and an edge exists if the two profiles are ``friends'' on Facebook. Metadata gives the social context for the relationship between ego and alter. For this analysis, only the largest connected component of the network was considered. This network has $N=329$ nodes and $E=1954$ edges.
\subparagraph{Within-organization Facebook friendships (\href{https://networks.skewed.de/net/facebook_organizations}{\texttt{facebook\_organizations}})~\cite{1303.3741v2}:} Six networks of friendships among users on Facebook who indicated employment at one of the target corporation. Companies range in size from small to large. Only edges between employees at the same company are included in a given snapshot. Node metadata gives listed job-type on the user's page. The particular networks named \texttt{S1}, \texttt{S2} have been used, with number of nodes and edges, $(N,E)$, given by (320, 2369), (165, 726), respectively.
\subparagraph{Little Rock Lake food web (\href{https://networks.skewed.de/net/foodweb_little_rock}{\texttt{foodweb\_little\_rock}})~\cite{Martinez_1991}:} A food web among the species found in Little Rock Lake in Wisconsin. Nodes are taxa (like species), either autotrophs, herbivores, carnivores or decomposers.  Edges represent feeding (nutrient transfer) of one taxon on another. For this analysis, a symmetrized version of the original directed network has been used. This network has $N=183$ nodes and $E=2494$ edges.
\subparagraph{NCAA college football 2000 (\href{https://networks.skewed.de/net/football}{\texttt{football}})~\cite{Girvan_2002}:} A network of American football games between Division IA colleges during regular season Fall 2000. This network has $N=115$ nodes and $E=613$ edges.
\subparagraph{Game of Thrones coappearances (\href{https://networks.skewed.de/net/game_thrones}{\texttt{game\_thrones}})~\cite{Beveridge_2016}:} Network of coappearances of characters in the Game of Thrones series, by George R. R. Martin, and in particular coappearances in the book ``A Storm of Swords.'' Nodes are unique characters, and edges are weighted by the number of times the two characters' names appeared within 15 words of each other in the text. This network has $N=107$ nodes and $E=352$ edges.
\subparagraph{Google+ (\href{https://networks.skewed.de/net/google_plus}{\texttt{google\_plus}})~\cite{Fire_2013}:} Snapshot of connections among users of Google+, collected in 2012. Nodes are users and a directed edge $(i,j)$ represents user $i$ added user $j$ to $i$'s circle. For this analysis, a symmetrized version of the original directed network has been used, considering only its largest connected component. This network has $N=201949$ nodes and $E=1496936$ edges.
\subparagraph{Jazz collaboration network (\href{https://networks.skewed.de/net/jazz_collab}{\texttt{jazz\_collab}})~\cite{GLEISER_2003}:} The network of collaborations among jazz musicians, and among jazz bands, extracted from The Red Hot Jazz Archive digital database, covering bands that performed between 1912 and 1940. This network has $N=198$ nodes and $E=2742$ edges.
\subparagraph{Zachary Karate Club (\href{https://networks.skewed.de/net/karate}{\texttt{karate}})~\cite{Zachary_1977}:} Network of friendships among members of a university karate club. Includes metadata for faction membership after a social partition. Note: there are two versions of this network, one with 77 edges and one with 78, due to an ambiguous typo in the original study. (The most commonly used is the one with 78 edges.). The particular network named \texttt{78} has been used. This network has $N=34$ nodes and $E=78$ edges.
\subparagraph{Les Misérables coappearances (\href{https://networks.skewed.de/net/lesmis}{\texttt{lesmis}})~\cite{knuth1993stanford}:} The network of scene coappearances of characters in Victor Hugo's novel ``Les Miserables.'' Edge weights denote the number of such occurrences. This network has $N=77$ nodes and $E=254$ edges.
\subparagraph{Malaria var DBLa HVR networks (\href{https://networks.skewed.de/net/malaria_genes}{\texttt{malaria\_genes}})~\cite{Larremore_2013}:} Networks of recombinant antigen genes from the human malaria parasite \emph{P. falciparum}. Each of the 9 networks shares the same set of vertices but has different edges, corresponding to the 9 highly variable regions (HVRs) in the DBLa domain of the var protein. Nodes are var genes, and two genes are connected if they share a substring whose length is statistically significant. Metadata includes two types of node labels, both based on sequence structure around HVR6. For this analysis, only the largest connected component of the network was considered. The particular network named \texttt{HVR\_9} has been used. This network has $N=297$ nodes and $E=7562$ edges.
\subparagraph{Scientific collaborations in network science (\href{https://networks.skewed.de/net/netscience}{\texttt{netscience}})~\cite{Newman_2006}:} A coauthorship network among scientists working on network science, from 2006. This network is a one-mode projection from the bipartite graph of authors and their scientific publications. For this analysis, only the largest connected component of the network was considered. This network has $N=379$ nodes and $E=914$ edges.
\subparagraph{Physician trust network (\href{https://networks.skewed.de/net/physician_trust}{\texttt{physician\_trust}})~\cite{Coleman_1957}:} A network of trust relationships among physicians in four midwestern (USA) cities in 1966. Edge direction indicates that node $i$ trusts or asks for advice from node $j$. Each of the four components represent the network within a given city. For this analysis, a  symmetrized version of the original directed network has been used, considering only its largest connected component. This network has $N=117$ nodes and $E=542$ edges.
\subparagraph{Multilayer physicist collaborations (\href{https://networks.skewed.de/net/physics_collab}{\texttt{physics\_collab}})~\cite{1408.2925v1}:} Two multiplex networks of coauthorships among  the Pierre Auger Collaboration of physicists (2010-2012) and among researchers who have posted preprints on arXiv.org (all papers up to May 2014). Layers represent different categories of publication, and an edge's weight indicates the number of reports written by the authors. These layers are one-mode projections from the underlying author-paper bipartite network. For this analysis, only the largest connected component of the network was considered. The particular network named \texttt{pierreAuger} has been used. This network has $N=475$ nodes and $E=7090$ edges.
\subparagraph{Political books network (\href{https://networks.skewed.de/net/polbooks}{\texttt{polbooks}})~\cite{Pasternak_1970}:} A network of books about U.S. politics published close to the 2004 U.S. presidential election, and sold by Amazon.com. Edges between books represent frequent copurchasing of those books by the same buyers. The network was compiled by V. Krebs and is unpublished. This network has $N=105$ nodes and $E=441$ edges.
\subparagraph{High school temporal contacts (\href{https://networks.skewed.de/net/sp_high_school}{\texttt{sp\_high\_school}})~\cite{Mastrandrea_2015}:} These data sets correspond to the contacts and friendship relations between students in a high school in Marseilles, France, in December 2013, as measured through several techniques. For this analysis,  symmetrized versions of the original directed networks have been used, considering only their largest connected component. The particular networks named \texttt{diaries}, \texttt{survey}, \texttt{facebook} have been used, with number of nodes and edges, $(N,E)$, given by (120, 502), (128, 658), (156, 1437), respectively.
\subparagraph{Student cooperation (\href{https://networks.skewed.de/net/student_cooperation}{\texttt{student\_cooperation}})~\cite{Fire_2012}:} Network of cooperation among students in the "Computer and Network Security" course at Ben-Gurion University, in 2012. Nodes are students, and edges denote cooperation between students while doing their homework. The graph contains three types of links: Time, Computer, Partners. For this analysis, only the largest connected component of the network was considered. This network has $N=141$ nodes and $E=297$ edges.
\subparagraph{9-11 terrorist network (\href{https://networks.skewed.de/net/terrorists_911}{\texttt{terrorists\_911}})~\cite{Krebs_2002}:} Network of individuals and their known social associations, centered around the hijackers that carried out the September 11th, 2001 terrorist attacks. Associations extracted after-the-fact from public data. Metadata labels say which plane a person was on, if any, on 9/11. This network has $N=62$ nodes and $E=152$ edges.
\subparagraph{Madrid train bombing terrorists (\href{https://networks.skewed.de/net/train_terrorists}{\texttt{train\_terrorists}})~\cite{Hayes_2006}:} A network of associations among the terrorists involved in the 2004 Madrid train bombing, as reconstructed from press stories after-the-fact. Edge weights encode four levels of connection strength: friendships, ties to Al Qaeda and Osama Bin Laden, co-participants in wars, and co-participants in previous terrorist attacks. This network has $N=64$ nodes and $E=243$ edges.
\subparagraph{Email network (\href{https://networks.skewed.de/net/uni_email}{\texttt{uni\_email}})~\cite{Guimer__2003}:} A network representing the exchange of emails among members of the Rovira i Virgili University in Spain, in 2003. For this analysis, a symmetrized version of the original directed network has been used. This network has $N=1133$ nodes and $E=10903$ edges.